\documentclass{article}

\usepackage{arxiv}

\usepackage[utf8]{inputenc} 
\usepackage[T1]{fontenc}    

\usepackage{url}            
\usepackage{booktabs}       
\usepackage{amsfonts}       
\usepackage{nicefrac}       
\usepackage{microtype}      
\usepackage{lipsum}		
\usepackage{graphicx}
\usepackage{float}
\usepackage{amsthm}
\usepackage{physics}
\usepackage{bm}
\usepackage{arydshln}

\usepackage[colorlinks=true, linkcolor=blue, citecolor=blue, urlcolor=blue, pdfborder={0 0 0}]{hyperref}

\theoremstyle{definition}
\newtheorem{dfn}{Definition}[section]
\newtheorem{prop}{Proposition}

\newcommand{\supplementaryfigures}{
  \renewcommand{\thefigure}{S\arabic{figure}}
  \setcounter{figure}{0}
}
\newcommand{\supplementarytables}{
  \renewcommand{\thetable}{S\arabic{table}}
  \setcounter{table}{0}
}





\title{Self-organized institutions in evolutionary dynamical-systems game}

\author{Kenji Itao\\
	Computational Group Dynamics Collaboration Unit,\\
    RIKEN Center for Brain Science, \\
    2-1 Hirosawa, Wako, Saitama 351-0198, Japan.\\
	\texttt{kenji.itao@riken.jp} \\
	\And
	Kunihiko Kaneko\\
	The Niels Bohr Institute, \\
    University of Copenhagen, \\
    Blegdamsvej 17, Copenhagen, 2100-DK, Denmark.\\
	\texttt{kunihiko.kaneko@nbi.ku.dk} \\
}

\renewcommand{\shorttitle}{Self-organized institutions in evolutionary dynamical-systems game}

\hypersetup{
pdftitle={Self-organized institutions in evolutionary dynamical-systems game},
pdfauthor={Kenji Itao, Kunihiko kaneko},
pdfkeywords={social institutions, evolutionary game theory, dynamical-systems game, statistical physics},
}

\begin{document}
\maketitle

\begin{abstract}
	Social institutions are systems of shared norms and rules that regulate people's behaviors, often emerging without external enforcement. They provide criteria to distinguish cooperation from defection and establish rules to sustain cooperation, shaped through long-term trial and error. While principles for successful institutions have been proposed, the mechanisms underlying their emergence remain poorly understood. To address this, we introduce the evolutionary dynamical-systems game, a framework that couples game actions with environmental dynamics and explores the evolution of cognitive frameworks for decision-making. We analyze a minimal model of common-pool resource management, where resources grow naturally and are harvested. Players use decision-making functions to determine whether to harvest at each step, based on environmental and peer monitoring. As these functions evolve, players detect selfish harvesting and punish it by degrading the environment through harvesting. This process leads to the self-organization of norms that classify harvesting actions as cooperative, defective, or punitive. The emergent norms for ``cooperativeness'' and rules of punishment serve as institutions. The environmental and players' states converge to distinct modes characterized by limit-cycle attractors, representing temporal regularities in socio-ecological systems. These modes remain stable despite slight variations in individual decision-making, illustrating the stability of institutions. The evolutionary robustness of decision-making functions serves as a measure of the evolutionary favorability of institutions, highlighting the role of plasticity in responding to diverse opponents. This work introduces foundational concepts in evolutionary dynamical-systems games and elucidates the mechanisms underlying the self-organization of social institutions by modeling the interplay between ecological dynamics and human decision-making.
\end{abstract}

\keywords{social institutions \and evolutionary game theory, \and dynamical-systems game \and statistical physics}

\section*{Introduction}
Social actions are influenced by the past, present, and anticipated behaviors of others, governed by institutions—shared norms, rules, and expectations that transcend individual agency \cite{greif1998historical, north1981structure, north2005understanding}. These institutions establish social regularities, guiding collective behavior and fostering order.
Even when selfish actions, such as the overuse of common-pool resources, offer individual benefits, people may refrain from it due to rules whose violation leads to future punishment. Indigenous institutions, in particular, have been shown to achieve sustainable resource use even in the absence of private property or centralized governance \cite{greif1998historical, ostrom1990governing, ostrom1994rules, haller2002common, garnett2018spatial}.
How, then, do such institutions arise without external enforcement? To address this, we propose a minimal model of common-pool resource management, illustrating the self-organization of social institutions.

For common-pool resource management, including forestry, fisheries, irrigation, and grazing, social institutions are essential to prevent overuse and avert the tragedy of the commons \cite{hardin1968tragedy, ostrom1990governing, cox2010exploring, cox2010review}. Several principles for successful management have been proposed, 
such as excluding outsiders and monitoring both resource conditions and users' behavior \cite{ostrom1990governing, cox2010review}. 
The criteria for acceptable resource use are typically shared within communities and vary depending on environmental and user conditions. For instance, resources may be allocated proportionally by land area in times of abundance, while minimum per capita allocations are prioritized during scarcity \cite{guillet1992covering}. Timing also plays a critical role; examples include restricting fishing to even-numbered years \cite{bayliss2010managing}, permitting farming only on designated days of the week \cite{bayliss2010managing, osei2017taboos}, or implementing turn-taking in water use \cite{geertz1980negara, ostrom1990governing, komakech2011understanding, bues2011agricultural}.
The distinction between cooperation and defection is not determined by the actions themselves but by the context in which they are performed. Identical actions may be viewed as cooperative or defective depending on situational factors, with criteria for ``cooperativeness'' formed over time through trial and error \cite{ostrom1990governing}.

The role of punishment in fostering cooperation has been widely studied in standard game theory. Social dilemmas are often modeled using the prisoner’s dilemma to examine conditions for the evolution of cooperation \cite{carrozzo2021tragedy}. However, these frameworks assume a fixed environment (represented by a payoff matrix) and prescribe criteria for cooperation and defection. To explain the origin of institutions, it is essential to explore how such criteria evolve to control environments. Empirical evidence suggests that harvesting is more likely to be accepted when the resources are abundant and the user is poor, whereas preservation is expected otherwise. \cite{ostrom1990governing, cox2010review}. This requests to couple decision-making in game theory with environmental dynamics, allowing decisions to adapt to changing environments.

In common-pool resource management, iterated actions (e.g., resource consumption) affect the environment (e.g., resource availability) \cite{tilman2020evolutionary, farahbakhsh2022modelling}. Thus, temporal changes in payoff matrices are driven by players’ actions. Previous studies have explored transitions between multiple payoff matrices in well-known models like the prisoner’s dilemma and the hawk-dove game, based on the history of actions or mutations \cite{tilman2020evolutionary, weitz2016oscillating, sadekar2024evolutionary, ito2024complete}. However, these transitions rely on ad hoc rules, leaving their underlying mechanisms unclear.
While another study has modeled institutions as second-order cooperation fostering first-order cooperation in social dilemmas, the origin of such coupling remains unexplained \cite{lie2024social}. This requests a natural setting where payoff matrices change directly in response to both environmental dynamics and players’ actions.

Individual decision-making is influenced by group-level institutions, which are themselves shaped by individuals’ decisions \cite{north1981structure, north2005understanding}. Human cognitive frameworks with bounded rationality are essential in the operation of institutions \cite{dequech2001bounded}. Cognitive scientists suggest that evolved human cognition, despite relying on incomplete information, optimizes contextual actions and can be ``better than rational'' \cite{cosmides1994better, boyer2012naturalness}.
To understand the endogenous changes in institutions—where individuals gradually modify existing ones—it is requested to show how the evolution of cognitive frameworks that guide judgments in diverse contexts shape institutions \cite{north1981structure, north2005understanding}.

These requests highlight two forms of interdependence to be addressed: between environmental dynamics and decision-making, and between individual decision-making and social institutions. Dynamical-systems games, where the payoff matrix changes based on players’ actions, provide a relevant framework for this purpose \cite{akiyama2000dynamical, akiyama2002dynamical, goldenfeld2011life}. These games have examined the tragedy of the commons where the optimal strategy is to wait for resources to grow, while individuals are incentivized to harvest prematurely for personal gain. While revealing phenomena like the evolution of coordinated harvesting, previous studies have insufficiently explained the underlying mechanisms or the conditions for their emergence due to model complexity. Moreover, the potential for developing game-theoretic concepts remains largely unexplored.

In this paper, we introduce a minimal model of an evolutionary dynamical-systems game that incorporates temporal changes in payoff matrices as functions of ecologically grounded environmental dynamics and players' actions. By examining the evolution of decision-making functions, we demonstrate the self-organization of social institutions for common-pool resource management. Unlike traditional approaches, our framework emphasizes the evolution of decision-making functions rather than behaviors. Then, we propose novel game-theoretic concepts relevant to institutions.

We introduce fundamental concepts in evolutionary dynamical-systems games. Modes are defined as typical game dynamics with temporal regularities in socio-ecological systems. These modes remain unchanged despite slight variations in players’ decision-making functions but transition discretely when these functions cross thresholds. Such regularities arise from norms for ``cooperativeness'' and potential punishment for violators, which collectively serve as social institutions. To evaluate the evolutionary favorability of these institutions, we propose the concept of evolutionary robustness of decision-making functions. Together, these concepts provide a framework to understand how social institutions self-organize and support the sustainable management of common-pool resources.

In the next section, we outline the general framework of the evolutionary dynamical-systems game and its relevance to social institutions. We then introduce a two-player harvesting game model. This model demonstrates the evolution of institutions facilitated by cognitive frameworks that distinguish acceptable and unacceptable behaviors. Finally, we define the above concepts mathematically and examine the mechanisms underlying the self-organization of social institutions.

\section*{Model}
First, we present the general framework of evolutionary dynamical-systems games, by following the previous work while refining key terminology \cite{akiyama2000dynamical, akiyama2002dynamical}. Subsequently, we introduce a minimal model by simplifying the previous model.

\subsection*{General framework of evolutionary dynamical-systems game}
The model includes environmental resources and players. The states $\bm{x} = (x_1, x_2, \cdots)$ represent the resource quantities, while $\bm{y} = (y_1, y_2, \cdots)$ represent players' richness. These states determine the payoffs for players' actions. Each player $i$ employs a decision-making function $f_i$ as their strategy, characterized by a set of parameters.

In each ``game step,'' players decide their action as \(a_i = f_i(\bm{x}, \bm{y})\), meaning that they monitor the states of both the environment and players for decision-making. These states change based on players' actions. Formally, this is expressed as \((\bm{x}(t + 1), \bm{y}(t + 1)) = g((\bm{x}(t), \bm{y}(t), \bm{a}(t)))\), where \(g\) represents the entire dynamics of the socio-ecological system.

The above game step is repeated for $T$ iterations in each generation. This sequence of game steps, defined by $\{(\bm{x}(t), \bm{y}(t), \bm{a}(t))\}_{t = 0}^T$, is called the ``dynamical-systems game.'' The temporal change of states $(\bm{x}(t), \bm{y}(t))$ as a function of states and actions makes it a dynamical system.

In each generation of the dynamical-systems game, the fitness of each player $i$ is calculated based on the sequence of their states $\{y_i(t)\}_{t = 0}^T$. Players then produce offspring in proportion to their fitness, with the offspring inheriting the decision-making function \(f\) with slight mutations, i.e., small changes in the parameters of $f$. Through this process, \(f\) evolves. The repeated iteration of generations is called the ``evolutionary dynamical-systems game.''

In the dynamical-systems game, players encounter various situations with different states $(\bm{x}, \bm{y})$, requiring a general cognitive framework $f$. Players' actions not only influence the states of the environment and players but also affect the future actions of others. If $f$ evolves effectively, it enables players to recognize selfish behavior and ``punish'' it by changing actions accordingly. Notably, the behavior identified as a ``violation'' is not prescribed; rather, the rules that govern actions emerge from the set of $f$ within the population, which evolves and is exogenous to any individual player. In this way, the set of $f$ establishes shared norms that serve as institutions \cite{greif1998historical}. The optimal decision-making function $f_i$ for a player $i$ depends on the $f$ of other players. Thus, by examining the interactions among these decision-making functions, the endogenous evolution of institutions is demonstrated.

\subsection*{Minimal model: Two players' harvesting game}
As a minimal model of common-pool resource management, we analyze a case where two players share a naturally growing resource  (e.g., fishery sites, forests, or grazing areas), modeling the tragedy of the commons. Here, harvesting offers immediate benefits, but delaying harvest enables the resource to grow and yield greater returns.

\begin{figure}[tb]
  \centering
   \includegraphics[width=\linewidth]{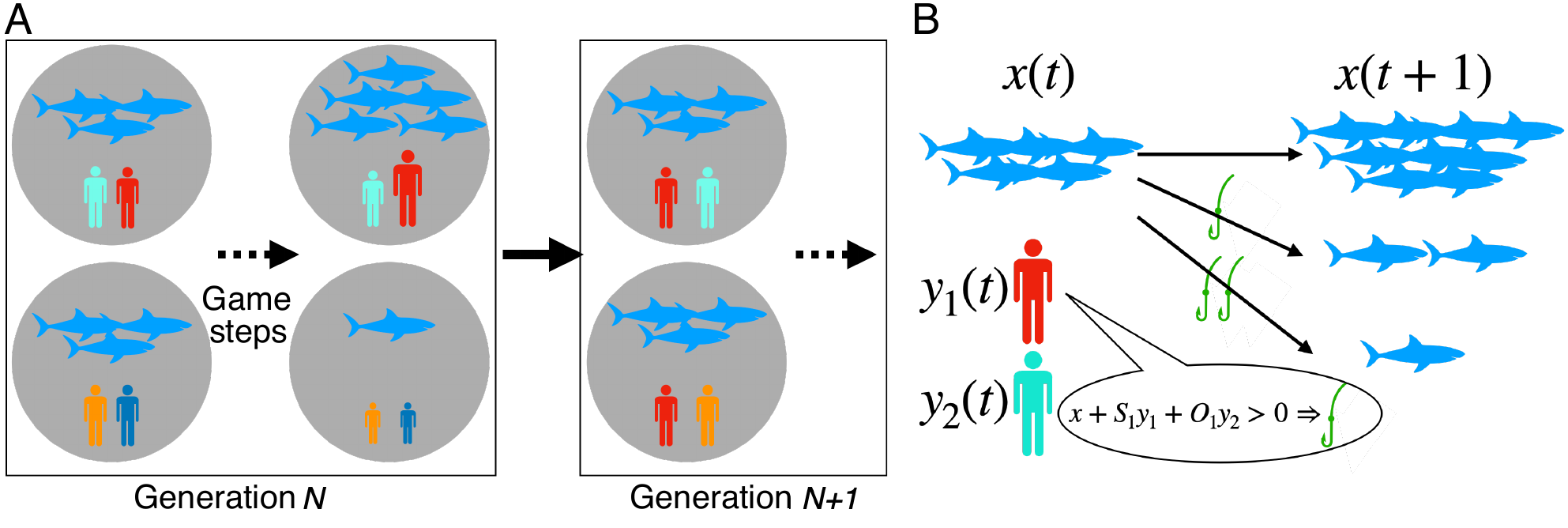}
   \caption{Schematic of the model. (A) Evolutionary dynamical-systems game. $N / 2$ pairs of players manage shared resources. Each pair iterates $T$ game steps, and players produce offspring who inherit their decision-making functions $f$. Dashed arrows indicate game steps, while solid arrows represent generation alternations. (B) A game step. Resources grow naturally, and players decide their harvesting actions based on their $f$ by observing the states of the environment and players.}
    \label{fig:DS_game_model_scheme}
\end{figure}

In the model schematically illustrated in Fig. \ref{fig:DS_game_model_scheme}, $N$ players are randomly matched into pairs. Each pair iteratively plays the game for $T$ steps. The observable states include the resource amount, $x(t)$, and the richness of both players (the self and opponent), $y_1(t)$ and $y_2(t)$. At each game step, the resource grows naturally according to $r(x) = x + \alpha(x - x^2)$, where $\alpha (> 0)$ represents the growth rate. This function indicates that the resource exhibits nonlinear growth when scarce but eventually converges to $x = 1$. In the following analysis, we set $\alpha = 1$, yielding $r(x) = 2x - x^2$.

The decision-making function $f$ is defined as a simple perceptron with parameterized weights. Player $i$'s action $a_i$ is determined by the function 
\begin{equation}
    a_i = \chi(x(t) + S_i y_\text{self}(t) + O_i y_\text{opponent}(t)),
\end{equation}
where the function $\chi(z)$ takes $\chi(z) = 1$ if $z > 0$, and $\chi(z) = 0$ otherwise. Here, $a_i = 1$ indicates harvesting, and $a_i = 0$ indicates waiting. Decision parameters $(S_i, O_i)$ characterize the player's ``strategy,'' where each parameter represents how the harvest decision becomes more likely with the self-state ($S$), and opponent's state ($O$). Thus, players decide to harvest by monitoring the richness of the environment and players.

The resource amount after harvesting is given by
\begin{equation}
    x(t + 1) = r(x(t))(1 - \beta)^{(a_1 + a_2)},
\end{equation}
where players harvest a fraction $\beta$ of the resource, reducing $x$ accordingly\footnote{Fishing methods such as net fishing can be represented in this manner, where a fixed fraction of the resource is harvested. Alternatively, a fixed amount may be harvested each time. However, the results discussed below remain largely unchanged in such cases as shown in Fig. S7.}. Here, we set $\beta = 2 / 3$, with $\beta < 1$ being a necessary condition for sustainability. The harvested resource is equally divided among the harvesters. Hence, the resource amount in the next step and the payoffs for the players are:
\begin{align}
   & (x(t+1), p_1(t), p_2(t)) 
   =
\begin{cases}
    \left(\frac{r(t)}{9}, \frac{4}{9}r(t), \frac{4}{9}r(t)\right) & \quad \text{if } a_1 = a_2 = 1, \\
    \left(\frac{r(t)}{3}, \frac{2}{3}r(t), 0\right) & \quad \text{if } a_1 = 1,\ a_2 = 0, \\
    \left(\frac{r(t)}{3}, 0, \frac{2}{3}r(t)\right) & \quad \text{if } a_1 = 0,\ a_2 = 1, \\
    \left(r(t), 0, 0\right) & \quad \text{if } a_1 = a_2 = 0.
\end{cases}
\end{align}
Hence, players earn more by harvesting alone than by harvesting simultaneously, making labor division advantageous.

The state change of a player due to harvesting is represented by
$y_i(t + 1) = (1 - \kappa) y_i(t) + p_i,$
where $\kappa$ is set to $0.2$. This parameter $\kappa$ represents the decay (or consumption) rate of a player's richness per step. Specifically, it indicates that players become poorer over time without new harvests, with their loss being proportional to their current richness.

The generality of the model is supported by its robustness to variations in parameter values and functional forms. Specifically, the qualitative results remain consistent even when different values of $\alpha$, $\beta$, and $\kappa$ are employed, or when alternative functional forms are used. For instance, the decision-making function may include additional terms as $a_i = \chi(E_i x + S_i y_{\text{self}} + O_i y_{\text{opponent}} + C_i)$, the growth function may take the form $r(x) = \min(\alpha x, 1.0)$, and the harvesting function may be modified to $x(t + 1) = \max(0.01, r(x(t)) - \beta(a_1 + a_2))$. These variations, as shown in Figs. S1-S7, do not affect the qualitative dynamics of the model.

\begin{table}[tb]
    \caption{Elements in the model. Variables change by game steps, evolvables change over generations, and parameters are fixed through evolution.}
  \label{table:DS_game_params}
   \centering
    \begin{tabular}{lrl} 
Sign & category & explanation \\\hline
$x$ & variable & The state of the environment \\
$y$ & variable & The state of the player \\
$a$ & variable & The action of the player \\
$h$ & variable & The fitness of the player \\\hdashline
$f$ & evolvable & Decision-making function of players \\
$S$ & evolvable & Weight for the self state in $f$ \\
$O$ & evolvable & Weight for the opponent's state in $f$ \\\hdashline
$\alpha$ & parameter & Resource growth rate (fixed at $1$) \\
$\beta$ & parameter & Harvesting fraction (fixed at $2/3$) \\
$\kappa$ & parameter & Decay rate of the state (fixed at $0.2$) \\
$N$ & parameter & The number of players in the system \\
$\mu$ & parameter & Mutation rate in the transmission of $f$
    \end{tabular}
\end{table}

After $T$ iterations, the fitness of player $i$ is calculated as 
$h_i = \sum_{t = 0}^{T} y_i(t) / T$.
Once all pairs have completed the game, player $i$ produces Poisson($h_i / \ev{h}$) offspring, where $\ev{h}$ is the average fitness of the population. Dividing by $\ev{h}$ ensures that the population size remains approximately $N$. Offspring inherit their parent's decision parameters $(S, O)$ with a small noise added to each parameter, drawn from a distribution with mean $0$ and variance $\mu^2$. In the next generation, the population is randomly matched into pairs and assigned to new resource sites (Fig. \ref{fig:DS_game_model_scheme}(A)).

Considering a system of $N (\gg 2)$ individuals, i.e., $N / 2$ pairs in each generation, players benefit evolutionarily by achieving higher fitness relative to other pairs, rather than merely outperforming their direct opponents. This introduces a form of multilevel evolution, enabling the analysis of macroscopic properties at the societal level \cite{hogeweg1994multilevel, spencer2001multilevel, henrich2006cooperation, traulsen2006evolution, itao2020evolution, itao2021evolution, itao2022emergence, itao2024formation}.

The initial conditions of the decision parameters are set to $S = O = 0$. In each generation, the game steps are iterated for $T = 1000$ steps, with the initial conditions are $x(0) = 0.1$, $y_1(0) = 1 + \eta_1$, and $y_2(0) = 1 + \eta_2$, where $\eta_1$ and $\eta_2$ are small noises drawn from a distribution with mean $0$ and variance $0.1$. 
The elements of the model are summarized in Table \ref{table:DS_game_params}.

\section*{Results}
\subsection*{Self-organization of social institutions}
\begin{figure*}[tb]
  \centering
   \includegraphics[width=\linewidth]{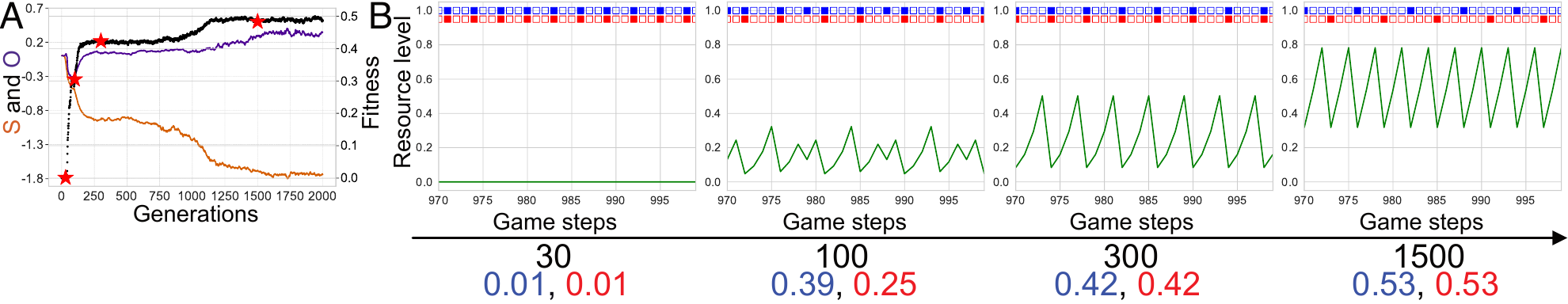}
   \caption{Example of the evolutionary dynamical-systems game. (A) Generational changes in the average values of the decision parameters $S$ (orange) and $O$ (purple). Black dots indicate the average fitness values (right axis). Red stars indicate the samples in (B). (B) Examples of game dynamics during the last 30 game steps across several generations. In each panel, blue (and red) boxes represent the harvesting actions of players 1 (and 2), while green represents the resource amount. The numbers below each panel indicate the generation (black) and the fitness values of players 1 (blue) and 2 (red). Parameters are set to $N = 300$ and $\mu = 0.03$.}
    \label{fig:DS_game_temporal_original}
\end{figure*}

Through evolution, periodic harvesting behaviors emerged, as shown in Fig. \ref{fig:DS_game_temporal_original}. In each generation, the state of the environment $x$ and the states of players $y_1$ and $y_2$ change over game steps based on the players' actions. As shown in Fig. \ref{fig:DS_game_temporal_original}(B), the temporal changes in $x$, $y_1$, and $y_2$ converge to periodic patterns after sufficient iterations. As long as the decision-making functions, or strategies, $f_1$ and $f_2$, remain unchanged, the same responses $a_1$ and $a_2$ occur for given values of $x$, $y_1$, and $y_2$. Consequently, the temporal dynamics of $(x(t), y_1(t), y_2(t))$ converge to an attractor. 
As there are no chaotic attractors in the present dynamics of $x(t)$,
\footnote{This model agrees with the standard logistic map $z=\lambda z(1-z)$, by the transformation $z= \alpha x / (1+\alpha)$, and $\lambda=1+\alpha$.
Hence, there is no chaos for $\alpha < 2.56995$.}, 
the attractors are limit cycles or fixed points, the latter being observed under overharvest conditions where $x(t) \to 0$.

The emergent limit cycles exhibited varying periods (i.e., harvesting frequencies) and synchronicity (synchronization or labor division). Lower harvesting frequency and synchronicity result in greater average resource amounts and higher fitness levels.
As generations progress, overharvesting initially occurs (generation 30 in Fig. \ref{fig:DS_game_temporal_original}(B)). Subsequently, players reduce their harvesting frequency, allowing the resource to recover. Initially, however, such ``cooperators'' are exploited by selfish players (generation 100). Over time, harvesting frequencies between players equalize, leading to coordination with synchronized harvesting (generation 300). Eventually, harvesting frequency decreases further, and labor is divided, with players alternating harvests (generation 1500). 
Throughout this process, the presence of $N / 2$ pairs makes achieving higher fitness relative to other pairs, rather than solely outperforming the opponent within the same pair, adaptive.

This evolution is driven by changes in the decision parameters of $f$, as shown in Fig. \ref{fig:DS_game_temporal_original}(A). Recall that player $i$ harvests if $x + S_i y_{\text{self}} + O_i y_{\text{opponent}} > 0$. Initially, the weights $S$ and $O$ decrease, causing players to harvest only when they are poor and the environment is rich, thereby reducing harvesting frequency.
At first, $S \simeq O$ indicates that players treat their self- and opponent's states as indistinguishable. Thus, players wait when either their self- or opponent's state is rich, allowing exploitation where one player accumulates wealth while the other does not. By reducing $S$ sufficiently below $O$, players adjust their strategy to wait when their self-states are rich but harvest when their opponents are solely rich. This punitive harvesting degrades the environment and lowers the fitness of selfish opponents, albeit at a cost to the punishers' own fitness.
At this stage, pairs without punishment or exploitation achieve higher fitness. Consequently, both players align their harvesting frequencies and harvest within acceptable ranges of each other. Further reductions in $S$ allow players to alternate harvesting, leading to the evolution of labor division.

The decision-making function $f$ determines the timing of harvesting actions. Evolved functions establish criteria for both sustainable and punitive harvesting. When players harvest sustainably, they perceive their opponents' actions as cooperative. Conversely, during punitive harvesting, they identify the opponent's defection. Thus, the criteria for ``cooperativeness'' are self-organized.
When both players' actions are symmetric and they regard each other's actions as cooperative, they adhere to shared ``norms.'' In this sense, social institutions are self-organized through the interaction of decision-making functions.

Notably, throughout the evolution, the variance in decision parameters $(S, O)$ was small, without significant differentiation, as shown in Fig. S8. This suggests that the emergent norms are widely shared across the evolved population.

Fig. S9 highlights the necessity of monitoring by presenting results where one of the decision parameters is fixed to $0$, making certain information invisible. 
We consider the model $f: \chi(Ex(t) + S_i y_\text{self}(t) + O_i y_\text{opponent}(t))$. When $E$ is fixed to $0$, coordinating behaviors evolve. However, when $S$ or $O$ is fixed to $0$, overharvesting becomes dominant. This indicates that information about players' states is essential in this evolution. Nevertheless, Fig. S9 also shows that observing players' actions, instead of their states, is sufficient.

\begin{figure}[tb]
  \centering
   \includegraphics[width=\linewidth]{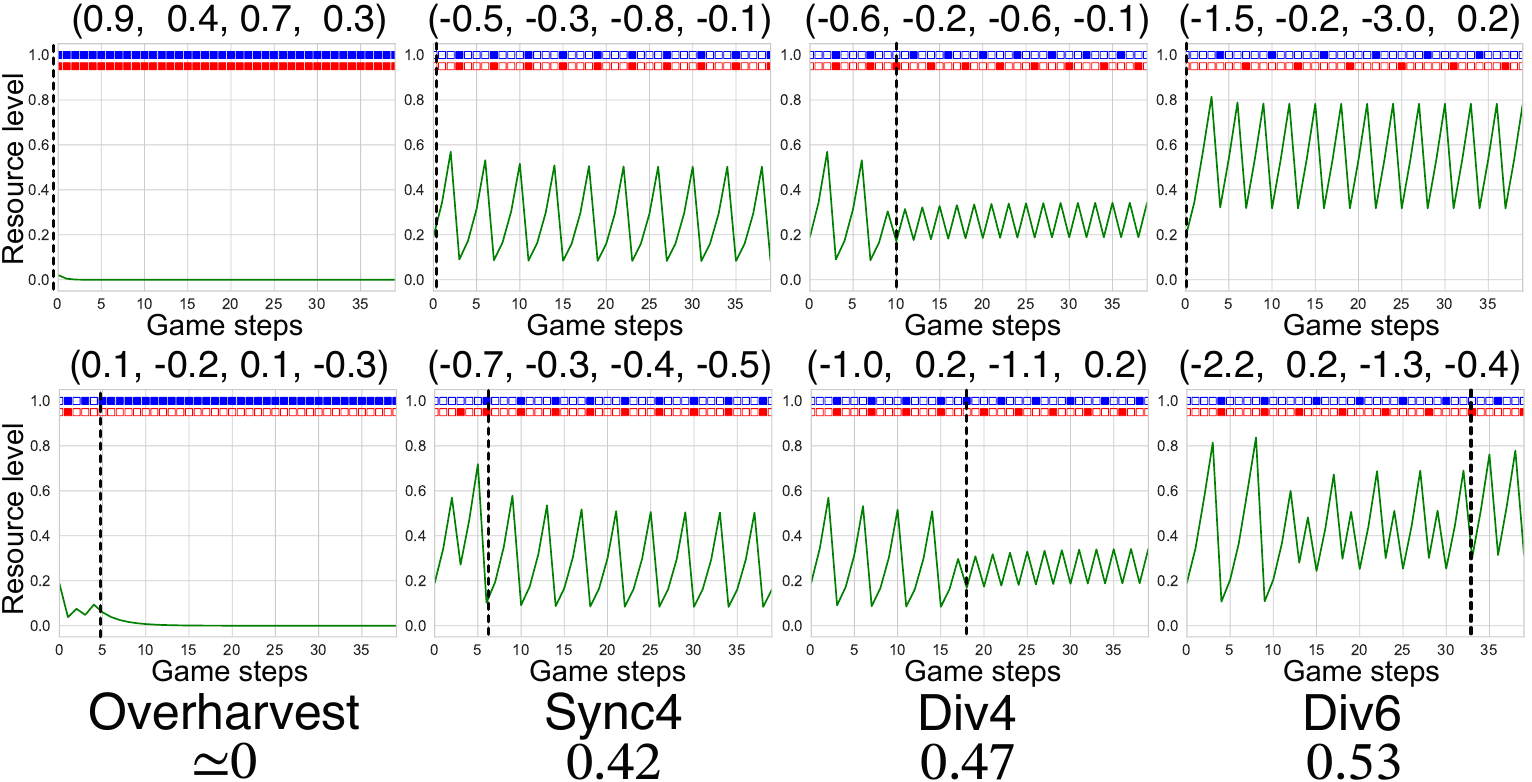}
   \caption{Modes of the game. In each panel, blue (and red) boxes represent the harvesting actions of players 1 (and 2), while green indicates the resource amount during the first 40 game steps. The top and bottom panels show results with different pairs of decision-making functions converging to identical modes (right side of the dashed line) after exhibiting different transient dynamics (left side). The numbers above each panel show $(S_1, O_1, S_2, O_2)$.
   Modes are classified based on the period and synchronicity of actions.  ``Sync'' (``Div'') indicates that the players' actions are synchronized (temporally divided), with the subsequent number denoting the action periods. "Overharvest" represents conditions where $x(t) \to 0$.
   The value at the bottom represents the fitness, which is identical for both players in each mode.}
    \label{fig:DS_game_phase_examples}
\end{figure}

\subsection*{Mode: Typical pattern in socio-ecological dynamics}
In Fig. \ref{fig:DS_game_temporal_original}(A), the fitness increases stepwise, forming several plateaus (e.g., generations 0–50, 200–800, and 1200–2000). Notably, during these plateaus, the periodic patterns remain identical even as the decision parameters change continuously across generations. Both the periods and phase shifts of actions, as well as the time series of $x$, $y_1$, and $y_2$ in the attractors, remain unchanged, resulting in identical fitness levels. Thus, each plateau corresponds to a specific periodic pattern.

Fig. \ref{fig:DS_game_phase_examples} illustrates typical examples of periodic patterns where the two players achieve identical fitness. Even with different pairs of decision-making functions (shown in the top and bottom panels), the games converge to identical limit cycles within 40 steps. In these examples, the two players either synchronize their harvesting actions or alternate in a half-phase shift with the same period. 

In some cases, however, the fitnesses of the two players are asymmetric, indicating exploitation. When this occurs, one strategy eventually dominates through evolution, causing the pattern to disappear over time. Fig. S10 shows examples with asymmetry and complexity.

To uncover the origins of these typical patterns in game dynamics, we mathematically analyze the dynamic processes governing the interdependent changes in actions and states. Generally, when the states change with a period $l$, the periodic pattern can be represented by the $3l$-dimensional vector of continuous states, $\{(x(i), y_1(i), y_2(i))\}_{i=1}^{l}$. However, Fig. \ref{fig:DS_game_phase_examples} suggests that these patterns depend solely on the action sequences of the two players, represented by the $2l$-dimensional vector of binary actions, $\{(a_1(i), a_2(i))\}_{i=1}^{l}$. 
Moreover, we confirmed that when the players' actions are periodic and occur with the same period, the periodic pattern is uniquely determined by two parameters: the period of the actions and the phase difference between the players.

Following these observations, we define the ``modes'' of periodic patterns as follows: 
\begin{dfn}
    Modes are defined as temporally periodic patterns of the states of the environment and players, $\{(x(i), y_1(i), y_2(i))\}_{i=1}^{l}$, where $l$ denotes the period.
\end{dfn}

\begin{prop}
    Modes depend solely on the action sequences $\{(a_1(i), a_2(i))\}_{i=1}^{l}$ (or their periods and synchronicity). Thus, modes remain invariant under changes in decision parameters, provided the action sequences are unchanged. For any given periodic action sequence, at most one mode exists.
\end{prop}

We provided the proof in the Supplementary text. Here, we briefly explain it using the simplest case, where players synchronously harvest once every $l$ steps. In this case\footnote{In general, $x$ converges to $x^\ast(l,\ \text{Sync}) = r^l(x^\ast(l,\ \text{Sync}))(1 - \beta)^2$, where $\beta$ denotes the harvesting fraction. The subsequent analysis remains qualitatively valid as long as $0 < \beta < 1$.}, $x$ after harvest converges to $x^\ast(l,\ \text{Sync}) = r^l(x^\ast(l,\ \text{Sync}))/9,$
where $r^l(x)$ represents the $l$-times iteration of $r(x)$. For a given $l$, at most one $x^\ast(l,\ \text{Sync}) > 0$ exists because $r^l(x)$ is a monotonically increasing and concave function of $x$.

Similarly, if players harvest once in an even-numbered period $l$ with a phase shift of $l/2$, $x$ after harvest converges to $x^\ast(l,\ \text{Div}) = r^{l/2}(x^\ast(l,\ \text{Div}))/3.$
A similar analysis can be applied to more complex modes with multiple harvests within a period. Therefore, the action sequences uniquely determine the time series of the states in the modes. Hereafter, modes are represented by their periods and synchronicity, such as ``Sync4'' and ``Div6.''

In the Supplementary text, we analytically calculated the value of $x^\ast$ and the time series of the states in each mode. Through this analysis, the fitness for each mode was identified, with the highest fitness observed in Div6.

The mode does not necessarily exist for any arbitrary action sequence. Consider the synchronized actions with $l = 2$ or $3$, i.e., $\{(a_1(t), a_2(t))\} = \{(1, 1), (0, 0), (1, 1), (0, 0), \cdots\}$ or $\{(1, 1), (0, 0), (0, 0), (1, 1), (0, 0), (0, 0), \cdots\}$. Since $r^l(x) / 9 < x$ for any $x > 0$, the equilibrium point $x^\ast > 0$ does not exist, and in the attractor, $x = y_1 = y_2 = 0$. Although $x \equiv 0$ satisfies $x = r^l(x / 9)$ for any $l$, at this point, the input to the decision-making functions remains constant, causing the action sequence to become $a \equiv 1$\footnote{$a \equiv x \equiv y_1 \equiv y_2 \equiv 0$ is an unstable attractor since $x = 0 + \delta$ ($\delta > 0$) increases over time. Hence, the stable overharvesting attractor is $a \equiv 1$ with $x \equiv y_1 \equiv y_2 \equiv 0$.}. Hence, periodic behaviors with $l = 2$ or $3$ are impossible.

By Prop. 1, even if the decision parameters change slightly, the mode remains identical as long as the action sequences are unchanged\footnote{Note that in dynamical systems, it is common for the period to remain unchanged even when parameters vary; however, the values of variables like $x$ in each attractor typically change. In the current model, if parameters such as $\alpha$ or $\beta$ change, the value of $x$ will also change. What is unique in this dynamical-systems game is that the values of $x$ and $y$ in the attractor remain completely unchanged despite slight changes in the parameters $S$ and $O$.}. 
As a result, in Fig. \ref{fig:DS_game_temporal_original}, fitness remains constant within some variations of $(S, O)$, but changes discontinuously beyond certain thresholds. The decision-making function determines the set of states $(x, y_1, y_2)$ where the player harvests. Although the boundary of this set gradually shifts with parameter variations, modes change only when the boundary crosses specific thresholds, altering action sequences and rendering the original mode unsustainable.

\subsection*{Iso-mode region: Parameter dependence of modes}
As the relationship between decision parameters and modes is discrete, it is essential to examine the regions in the decision parameter space where the mode remains unchanged and those where it transitions. To understand the parameter dependence of modes, we define the iso-mode region.
\begin{dfn}
    Iso-mode regions are defined as the connected sets of parameters $(S_1, O_1, S_2, O_2)$ within which the modes, and consequently the fitness, remain identical.
\end{dfn}

\begin{figure}[tb]
  \centering
   \includegraphics[width=0.6\linewidth]{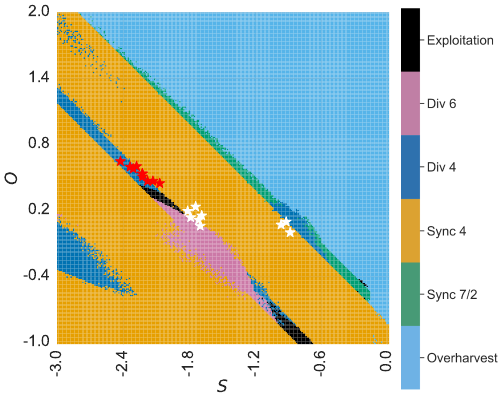}
   \caption{Iso-mode regions. The dominant mode for each $(S, O)$ is shown in different colors by considering a game with two players sharing the same $(S, O)$. (Here, classification is based on harvesting frequency and fitness. Thus, modes Div4 and Div8/2 are plotted in the same color.) White (and red) stars indicate average $(S, O)$ at generation $10,000$ for $N = 100$ and $\mu = 0.03$ ($N = 3000$ and $\mu = 0.3$). }
    \label{fig:DS_game_isomode}
\end{figure}

Considering iso-mode regions in 4D space is overly complex. Therefore, we examine the case where both players adopt identical strategies (i.e., playing the game with oneself) to illustrate the iso-mode regions in 2D space, as shown in Fig. \ref{fig:DS_game_isomode}. (Note that, after evolution, the variance in strategies among players is small, allowing this restriction to 2D space in most cases, as shown in Fig. S8). 
In general, larger values of $(S, O)$ lead to higher harvesting frequencies. Synchronized modes broadly emerge in regions satisfying $S + O \simeq \text{constant}$, regardless of the relative weights of $S$ and $O$, as self and opponent states are equal during synchronization. In contrast, maintaining labor division requires a specific balance between $S$ and $O$, which constrains the iso-mode regions.

\begin{prop}
    If $(S_1, O_1, S_1, O_1)$ and $(S_2, O_2, S_2, O_2)$ are in the same iso-mode region, then $(S_1, O_1, S_2, O_2)$ is also in that region.
\end{prop}

\textit{Proof.} By Prop. 1, the games for $(S_1, O_1, S_1, O_1)$ and $(S_2, O_2, S_2, O_2)$ yield identical time series of actions and states. Therefore, the decision-making functions $f_1: (S_1, O_1)$ and $f_2: (S_2, O_2)$ produce identical action sequences for the states in the mode. Consequently, the original mode is preserved in the game for $(S_1, O_1, S_2, O_2)$. 
(Additional evidence is provided in Figs. \ref{fig:DS_game_invadability}(A-F), as will be discussed later.)

Evolution converges to specific regions depending on the number of players $N$ and the mutation rate of decision parameters $\mu$, as shown in Fig. \ref{fig:DS_game_isomode}. It is reasonable that evolution converges to several fitter modes such as Div6 and Div4. 
Notably, when strategies within the same iso-mode region are played against each other, the resulting mode and fitness are identical. This suggests that iso-mode regions appear evolutionarily neutral, with each subset of iso-mode regions being equally accessible. However, Figs. \ref{fig:DS_game_isomode} demonstrate that only a limited subset of these regions is actually reached. Moreover, while Div4 regions exist in multiple locations, evolution realizes only a specific iso-mode region of Div4. 
This observation requires considering games involving strategies from different iso-mode regions, as such strategies can emerge through mutation during evolution.

\subsection*{Evolutionary robustness: conditions for dominance}
We sampled six strategies, namely Div4$a_1$, Div8/2$a_1$, Div6$a_1$, Div4$b_1$, Div4$c_1$, and Div6$a_2$. The terminology is based on three components: the mode resulting from the game played with itself (e.g., Div4), the identifier of the iso-mode region (e.g., $a$), and the identifier of the strategy within that region (e.g., $1$). 
For example, Div6$a_1$ and Div6$a_2$ are strategies from the same iso-mode region, and thus a game between them converges to Div6. In contrast, Div4$a_1$ and Div4$b_1$ are strategies from different iso-mode regions, so the game between them does not necessarily converge to Div4.

Div4$a_1$ and Div6$a_1$ evolve under conditions of $N = 100$ and $\mu = 0.03$, while Div8/2$a_1$ evolves under $N = 3000$ and $\mu = 0.3$, as shown in Fig. \ref{fig:DS_game_isomode}. In contrast, Div4$b_1$, Div4$c_1$, and Div6$a_2$ rarely evolve in our simulations. 
Note that Div8/2 exhibits a complex periodic behavior, where each player harvests twice within eight steps, as shown in Fig. S10. Its fitness is nearly identical to that of Div4 and is therefore plotted in the same color in Fig. \ref{fig:DS_game_isomode}.

\begin{figure}[tb]
  \centering
   \includegraphics[width=\linewidth]{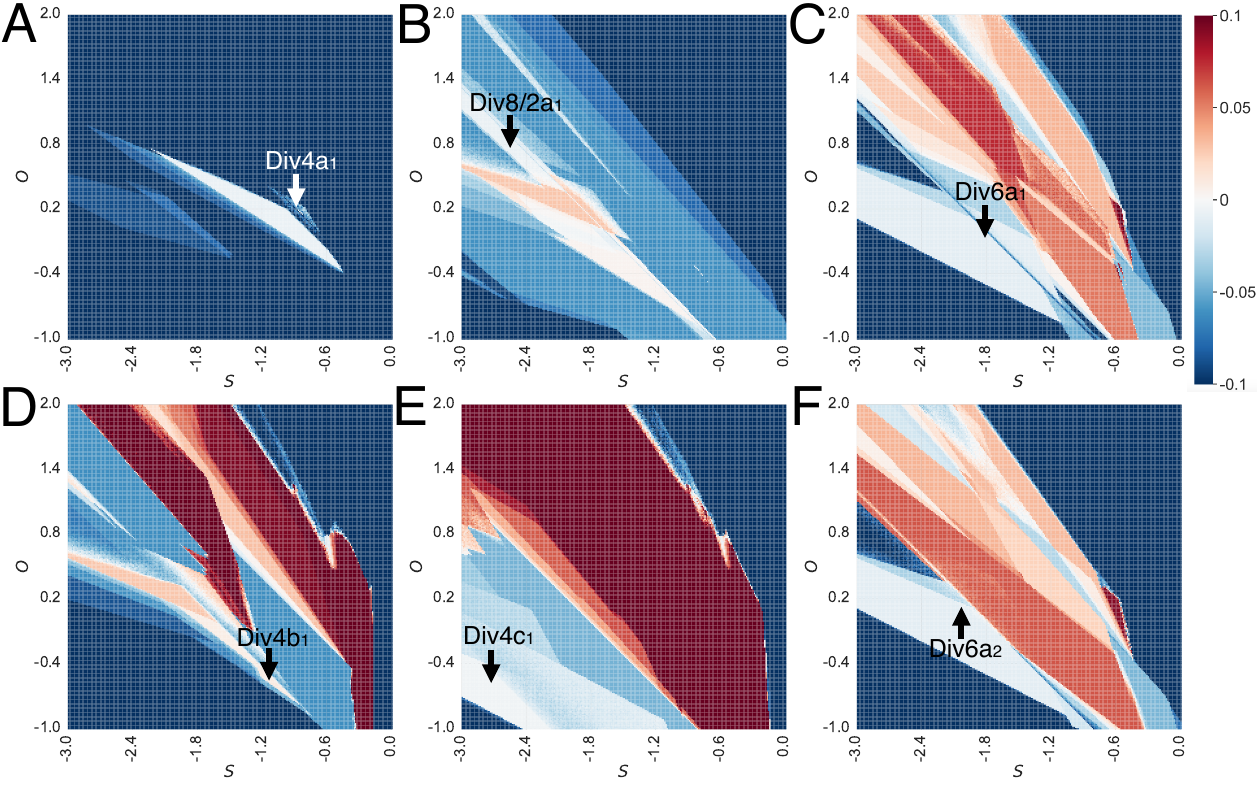}
   \caption{Evolutionary robustness of strategies (Div4$a_1$ (A), Div8/2$a_1$ (B), Div6$a_1$ (C), Div4$b_1$ (D), Div4$c_1$ (E), and Div6$a_2$ (F), named as ``[mode name][region ID][strategy number]''). One strategy is fixed at the arrowhead while the opponent's strategy is varied in parameter space. Colors indicate the opponents' fitness as the difference from the original mode. Fixed strategies can be invaded by those plotted in red.}
    \label{fig:DS_game_invadability}
\end{figure}

To assess the evolutionary favorability of these strategies, we fixed each strategy for one player while varying the opponent's strategy within 2D space. Figs. \ref{fig:DS_game_invadability}(A-F) illustrate the difference between the fitness of the opponent and that of the original mode.
Positive values (plotted in red) indicate that the fixed strategy can be invaded by those strategies. 
Within the region plotted in white, which includes the iso-mode region, the fitness difference is nearly $0$. This implies that the identical mode is maintained within iso-mode regions, supporting Prop. 2. In contrast, when two strategies from disjoint but same-colored regions (e.g., Div4$a_1$ and Div4$b_1$) play the game, one strategy can invade the other.

Figs. \ref{fig:DS_game_invadability}(C-F) illustrate that the strategies Div4$b_1$, Div4$c_1$, Div6$a_1$, and Div6$a_2$ are vulnerable to invasion by frequent harvesters, located in the upper-right region of the 2D space. By comparing Figs. \ref{fig:DS_game_invadability}(C, F), it is evident that although Div6$a_1$ and Div6$a_2$ are in the same iso-mode region, the mode of the game with distant neighbors differ significantly. 
Since Div6$a_1$ is slightly more robust against invasion, it evolves most frequently within the iso-mode region for Div6.

Notably, the Div4$a_1$ and Div8/2$a_1$ strategies are robust against invaders, as shown in Figs. \ref{fig:DS_game_invadability}(A, B). For Div8/2$a_1$, certain lower regions (plotted in pink) exhibit higher fitness than the fitness of the mode Div8/2. However, Div8/2$a_1$ still achieves higher fitness than its opponent (Fig. S11(B)). 
In other words, the fitness ordering is as follows: Div8/2$a_1$ against this opponent $>$ this opponent against Div8/2$a_1$ $>$ the fitness of Div8/2. Thus, Div8/2$a_1$ is evolutionarily robust across a broad range of regions, remaining the dominant strategy even while allowing minor invasions.

While it is reasonable to expect that modes with higher fitness evolve more frequently, Div6 rarely evolves under high mutation rates ($\mu$) despite being the fittest. Understanding the adaptability of Div8/2 requires examining games involving distinct strategies, where robustness against invasion plays a crucial role. The Div6$a_1$ strategy evolves only when $\mu$ is small, as it is highly vulnerable to invasion by frequent harvesters. In contrast, the Div8/2$a_1$ strategy dominates under higher $\mu$ due to its greater robustness.

Following these observations, we define $\epsilon$-evolutionary robustness for strategies.
\begin{dfn}
    A strategy $f^\ast$ is said to have $\epsilon$-evolutionary robustness if, for any decision-making function $f$ within an $\epsilon$-radius ($\epsilon > 0$) around $f^\ast$ in the 2D space of $(S, O)$, the fitness of $f$ and $f^\ast$ satisfy either of the following conditions:
    \begin{align*}
        \text{(i) } &h(f^\ast, f^\ast) \ge h(f, f^\ast), \text{ or} \\
        \text{(ii) } &h(f^\ast, f) \ge h(f, f^\ast) > h(f^\ast, f^\ast) \ge h(f, f),
    \end{align*}
    where $h(f_1, f_2)$ represents the fitness of $f_1$ in the game between $f_1$ and $f_2$. 
    The case of $\infty$-evolutionary robustness is termed global evolutionary robustness.
\end{dfn}

This definition implies that a strategy has $\epsilon$-evolutionary robustness when the best option for an opponent within the $\epsilon$-radius is either to adopt the original strategy (i) or to be exploited by it (ii).

Under condition (ii), although $f^\ast$ allows partial invasion by $f$, $f^\ast$ is guaranteed to remain the majority strategy. Since $h(f^\ast, f) > h(f, f)$ and $h(f, f^\ast) > h(f^\ast, f^\ast)$, $f^\ast$ can invade a population dominated by $f$ and vice versa. Notably, both $h(f^\ast, f)$ and $h(f, f^\ast)$ are larger than the fitness of the original mode $h(f^\ast, f^\ast)$. This occurs when the original mode is inefficient, allowing simultaneous improvement in both players' fitness, as observed in the case of Div8/2 in Fig. \ref{fig:DS_game_invadability}(B) and Fig. S11(B).
Furthermore, since $h(f^\ast, f^\ast) + h(f^\ast, f) \ge h(f, f) + h(f, f^\ast)$, the average fitness of $f^\ast$ exceeds that of $f$ when both strategies are equally frequent. Therefore, $f^\ast$ becomes the majority strategy.

The definition is inspired by the concept of an Evolutionarily Stable Strategy (ESS) but is weaker than ESS \cite{apaloo2009evolutionary}. Here, we adapt the concept to apply to dynamical-systems games, where strategies are characterized by continuous parameters while modes shift discretely. In this context, close neighbors are evolutionarily neutral due to Props. 1 and 2. Thus, no strategy is an ESS, necessitating the above definition.

\begin{prop}
    For any iso-mode region, strategies with $\epsilon$-evolutionary robustness exist.
\end{prop}

\textit{Proof.} By Prop. 2, $h(f^\ast, f^\ast) = h(f, f^\ast)$ for any $f$ within the same iso-mode region as $f^\ast$. Thus, any strategy has evolutionary robustness within its iso-mode region. Unless a strategy is located at the edge of the iso-mode region, it possesses $\epsilon$-evolutionary robustness for at least some finite $\epsilon$.

For example, within the range illustrated in Fig. \ref{fig:DS_game_invadability}, Div4$a_1$ and Div8/2$a_1$ exhibit global evolutionary robustness, while the other strategies have finite $\epsilon$-evolutionary robustness. This implies that they can be invaded by some strategies outside the $\epsilon$-radius. 
The approximate values of $\epsilon$ are as follows: 0.5 for Div6$a_1$, 0.5 for Div4$b_1$, 1.2 for Div4$c_1$, and 0.1 for Div6$a_2$, as summarized in Table \ref{tab:DS_game_strategy}.

\begin{figure}[tb]
  \centering
   \includegraphics[width=\linewidth]{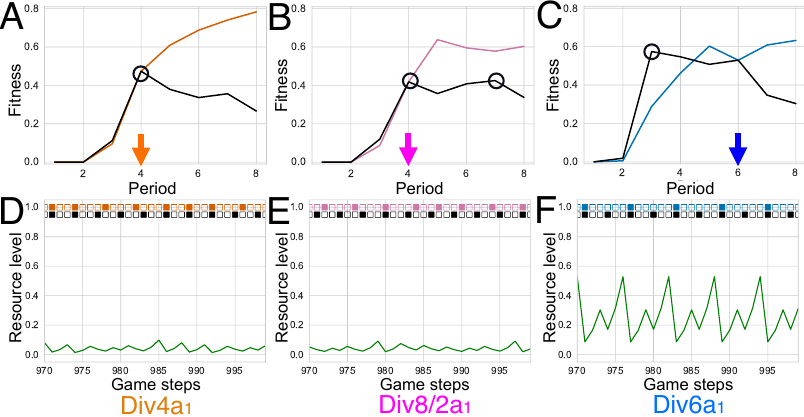}
    \caption{
    (A, B, C) Plasticity of Div4$a_1$ (A), Div8/2$a_1$ (B), and Div6$a_1$ (C) strategies. Here, the mechanical opponent harvests with a fixed period regardless of the state. The plots show the fitness of the strategy (orange for Div4$a_1$, pink for Div8/2$a_1$, and blue for Div6$a_1$) and the opponent (black). Black circles indicate the optimal period for the opponent, while colored arrows show the action periods of the strategies playing the game with themselves.
    (D, E, F) The game dynamics between the period-$3$ mechanical opponent and Div4$a_1$ (D), Div8/2$a_1$ (E), and Div6$a_1$ (F).}
    \label{fig:DS_game_stability}
\end{figure}

To understand the differences in evolutionary robustness, we analyze the response of strategies against distinct opponents. Fig. \ref{fig:DS_game_stability} illustrates the responses of the strategies Div4$a_1$, Div8/2$a_1$, and Div6$a_1$ against a ``mechanical'' opponent that harvests with a fixed period. 
As shown in Fig. \ref{fig:DS_game_stability}(A, B), for the Div4$a_1$ and Div8/2$a_1$ strategies, the optimal strategy for the opponent is to act with a period of $4$. This confers evolutionary robustness to these strategies. Fig. \ref{fig:DS_game_stability}(B) shows that the best choice for the opponent is either to adopt the original strategy\footnote{The mode Div8/2 exhibits a complex periodic pattern where players act twice in a period of $8$ with labor division. Such a pattern is infeasible against the mechanical opponent, and the resulting mode between Div8/2$a_1$ and the mechanical opponent with a period of $4$ becomes Sync4.} (period $4$) or to be exploited (period $7$). In contrast, the Div6$a_1$ strategy is vulnerable to exploitation by players harvesting with a period of $3$, as shown in Fig. \ref{fig:DS_game_stability}(C). Therefore, this strategy exhibits limited $\epsilon$-evolutionary robustness, applicable only up to a finite value of $\epsilon$.

Figs. \ref{fig:DS_game_stability}(D, E, F) show the modes of the game between each strategy and the period-$3$ mechanical opponent. In these cases, the Div4$a_1$ and Div8/2$a_1$ strategies adapt by harvesting more frequently than their original period of $4$, thereby preventing exploitation. However, Div6$a_1$ retains its original period-$6$ behavior even when the opponent harvests with a period of $3$, resulting in being exploited.
To achieve robustness against invasion by frequent harvesters, it is necessary to harvest more as a form of punishment. Thus, plasticity in decision-making—specifically, the ability to adjust harvesting frequency based on the opponent's strategy—is crucial to preventing invasion.

Additionally, Fig. \ref{fig:DS_game_stability}(A-C) suggests that fitness against less frequent harvesters varies across strategies. Among these, Div4$a_1$ exploits less frequent harvesters most effectively, due to its plasticity in accelerating harvesting frequency.
Plasticity serves two key roles: punishing frequent harvesters and exploiting less frequent harvesters. Therefore, high plasticity is critical for enhancing evolutionary favorability.

\subsection*{Evolutionary favorability of strategies}
Table \ref{tab:DS_game_strategy} summarizes the properties of the strategies in Fig. \ref{fig:DS_game_invadability}. Although Div6 is the fittest mode, the Div6$a_1$ strategy exhibits limited evolutionary robustness. In contrast, the less fit strategies Div8/2$a_1$ and Div4$a_1$ demonstrate global evolutionary robustness.

In addition to evaluating the fitness of the original mode and the radius of $\epsilon$-evolutionary robustness, we measured the average fitness against 100 opponents whose strategies were generated by introducing random variations to the fixed strategies. These variations were modeled as $(S+\eta_1, O+\eta_2)$, where $(S, O)$ represents the original strategy, and $\eta_1$ and $\eta_2$ are noise terms with a variance of $0.1$. This approach estimates the average fitness under conditions of strategy diversity in evolutionary scenarios.
The average fitness of the Div4$a_1$ strategy decreases when the opponent's strategy is distributed around itself. This occurs because its proximity to overharvesting strategies often leads to punitive overharvesting (see Fig. \ref{fig:DS_game_isomode}). While punishment is essential to prevent the invasion of selfish harvesters, excessive punishment also reduces the punisher’s own fitness.

\begin{table}[tb]
    \centering
    \caption{Properties of each strategy. The values represent the fitness of the original mode, the average fitness against opponents distributed around the strategy with the noise of variance $0.1$, and the radius of $\epsilon$-evolutionary robustness, respectively.}
    \label{tab:DS_game_strategy}
    \begin{tabular}{c|ccc}
       Strategy  & Original Fitness & Average Fitness & $\epsilon$  \\\hline
       Div4$a_1$  & 0.47 & 0.31 & $\infty$ \\
       Div8/2$a_1$ & 0.46 & 0.42 & $\infty$\\
       Div6$a_1$ & 0.53 & 0.51 & 0.5\\
       Div4$b_1$  & 0.47 & 0.45 & 0.5 \\
       Div4$c_1$ & 0.47 & 0.46 & 1.2\\
       Div6$a_2$ & 0.53 & 0.52 & 0.1
    \end{tabular}    
\end{table}

In evolutionary processes, strategies fluctuate within a society, making it fundamentally important to both maintain fitter modes with neighboring strategies in the decision parameter space and gain an advantage over distant strategies. The relative importance of these factors depends on the population size $N$ and the mutation rate $\mu$. Larger values of $N$ and $\mu$ increase the diversity of strategies, amplifying the importance of plasticity over fitness in interactions with players located closely in the decision parameter space.

This framework explains the evolution observed in Fig. \ref{fig:DS_game_isomode}. The Div6$a_1$ strategy dominates only under conditions of small $N$ and $\mu$ due to the lack of evolutionary robustness. Similarly, the evolution of the Div4$a_1$ strategy is limited to conditions with small $N$ and $\mu$, as it frequently opts for punitive overharvesting. 
For large $N$ and $\mu$, the Div8/2$a_1$ strategy evolves due to its higher plasticity. Even within iso-mode regions, the degree of plasticity against strategies outside the regions varies. Consequently, evolutionary processes often result in directional changes within iso-mode regions, ultimately converging to a specific subset of these regions.

\section*{Discussion}
\subsection*{Mechanism of the emergence of institutions}
By introducing an evolutionary dynamical-systems game framework, we demonstrated the mechanisms underlying the self-organization of social institutions. Our model captures the dynamics of common-pool resource management, where players’ actions are coupled with environmental dynamics. The evolution of decision-making functions—taking environmental and peer monitoring as inputs—leads to the emergence of social regularities. The self-organized periodic behavior arises through the development of shared norms that govern the acceptability of others’ actions and the punishment of violators. 
Thus, our model represents the interdependence between individual decision-making and social institutions, where individual decision-making operates within the constraints of existing institutions while simultaneously shaping future institutions.

Here, the norms for the criteria of ``cooperativeness'' and the punishment for violations were self-organized. Players decide whether to harvest in each game step based on decision-making functions. Through evolved functions, players detect selfish behavior in their opponents and punish it by harvesting to degrade the environment. 

Evolved decision-making functions determine whether a single harvesting action is perceived by the opponent as cooperation (if the opponent decides not to punish), defection (if the opponent decides to punish), or punishment (as a reaction to defection). These functions generate context-dependent norms of cooperativeness, defining acceptable harvesting actions of others. Such norms, shared within the evolved population and exogenous to individual players—since whether one's action is perceived as cooperative depends on others—represent the emergent institutions in our model. 
This approach overcomes the limitations of previous models, where the options for cooperation, defection, and punishment were prescribed \cite{boyd1992punishment, boyd2003evolution, chowdhury2021eco, carrozzo2021tragedy, lie2024social}.

By modeling the evolution of the cognitive framework, represented as decision-making functions, we showed the emergence of context-dependent norms, referencing environmental conditions and players' richness. Decision parameters (i.e., the parameters in the decision-making functions) evolved so that the likelihood of harvesting decreases with self-richness and increases with the opponent’s richness. 
This result indicates that resource use is perceived as acceptable when the user is poor and both the environment and the monitor are rich, consistent with institutional theories \cite{ostrom1990governing, cox2010review}.

Our analysis highlights the necessity of peer monitoring for the maintenance of institutions. In a model variant where information about self- or opponent's richness was unavailable, cooperation became unsustainable. However, when players' actions (rather than their richness) were observable, cooperation was achieved. 
This finding aligns with previous studies emphasizing the importance of monitoring for sustaining cooperation \cite{ostrom1990governing, cox2010review}.

\subsection*{General concepts in evolutionary dynamical-systems games}
The temporal changes in the states of the environment and players converge to ``modes'' of the games, i.e., limit-cycle attractors, including periodic synchronized harvesting and labor division. This occurs because the same actions are chosen for the same states as long as the decision-making functions remain unchanged. 
The specific form of the limit-cycle attractors depends on the resource growth rate $\alpha$ and the harvesting fraction $\beta$. However, the convergence to attractors, and thus the emergence of modes, is a universal feature of dynamical-systems games that couple decision-making with environmental dynamics. 
These modes represent the temporal regularities in socio-ecological systems enabled by emergent institutions, consistent with empirical observations of periodic harvesting (e.g., fisheries only in even-numbered years) and turn-taking (e.g., biweekly water use) \cite{bayliss2010managing, ostrom1990governing}.

Furthermore, we showed that identical modes emerge even when decision parameters differ, as long as the players' action sequences remain unchanged. In other words, only typical modes persist.
Players' actions remain stable unless decision parameters change beyond a threshold, causing different action sequences and leading to bifurcations in the dynamical system. Accordingly, we define the ``iso-mode region'' in the decision parameter space, where identical modes emerge. Within iso-mode regions, the game dynamics converge to the same modes, resulting in identical fitnesses. 
This iso-mode region is a generic feature of dynamical-systems games and provides a foundational basis for the stability of social institutions.

Finally, we introduced the concept of evolutionary robustness to explain which decision-making functions, and consequently institutions, prevail through evolution. Functions that sustain cooperation among similar strategies may face extinction if they are vulnerable to exploitation by selfish invaders. 
In games between functions within the same iso-mode region, all functions produce identical action sequences and fitness. However, when functions from the same iso-mode region interact with opponents from different iso-mode regions, their responses (i.e., action sequences and fitness) can diverge, leading to variations in robustness. Evolutionarily robust functions that maintain a majority against mutant strategies are favored.
This robustness is a universal concept in evolutionary dynamical-systems games, providing a formal basis for the evolutionary favorability of institutions and complementing economic efficiency as a criterion for institutional success.

Evolutionary robustness necessitates plasticity in decision-making. To prevent the invasion of selfish harvesters, players must punish them, which requires the ability to adjust harvesting frequency plastically. Therefore, a robust institution must maintain stability in preserving regularity when everyone adheres to the norms while also exhibiting plasticity in actions to punish norm violators. This balance is reminiscent of the Tit for Tat strategy in the prisoner’s dilemma game, where cooperation is sustained through reciprocal adjustments to the opponent’s behavior \cite{rapoport1965prisoners}.

\subsection*{Connections to institutional theory and broader implications}
The evolutionary dynamical-systems game illustrates the gradual process of institutional change, where individuals modify institutions within the constraints of pre-existing ones \cite{north1981structure, north2005understanding}. The decision-making functions in the population, which constitute institutions, determine the optimal functions selected within each generation to shape future institutions. This iterative renewal of decision-making functions drives the stepwise evolution of institutions.
This framework demonstrates the emergence of institutions, addressing longstanding calls in institutional theory \cite{ostrom1990governing, north2005understanding}.

Previous works have introduced transitions in payoff matrices over time \cite{weitz2016oscillating, sadekar2024evolutionary, ito2024complete}. However, the underlying processes driving such transitions remain unclear, as these transitions are often incorporated ad hoc. In our model, by coupling game dynamics with ecologically natural environmental dynamics, payoff matrices change naturally. 
The effective payoff matrices depend on the norms within the population (i.e., potential punishment through environmental degradation). These include the Prisoner's Dilemma game, which leads to overharvesting; the Stag Hunt game, which promotes synchronized coordination; and the Complementarity game, which fosters division of labor \cite{beheim2024cultural}. This framework provides a basis for discussing which institutions (and associated payoff matrices) can emerge, depending on environmental conditions and players' cognition.

Additionally, previous studies have emphasized indirect reciprocity, where reputations play a central role in explaining the evolution of large-scale cooperation \cite{leimar2001evolution, fujimoto2023evolutionary, murase2024indirect}. However, from the perspective of institutional economics, cooperation is primarily maintained through shared expectations that individuals will be punished for misconduct \cite{greif1998historical}. 
Our results demonstrate the emergence of a cognitive framework where individuals punish selfish players, leading to the evolution of coordinated behaviors driven by potential punishment. These self-organized norms reflect customary law, as highlighted in previous institutional theory \cite{ostrom1990governing}.

We have explored a minimal evolutionary dynamical-systems game, leaving several limitations to be addressed in future studies. First, extending the model to include multiple players and resources could provide insights into spontaneous group formation and resource allocation, potentially shedding light on the emergence of property rights \cite{andrews2024cultural, clark2024effects}. 
Second, the effects of varying initial conditions should be examined to reflect social-ecological feedback and the slow environmental changes induced by human activities in the Anthropocene \cite{meyfroidt2013environmental, clark2024effects, steffen2011anthropocene, lewis2015defining, ellis2024anthropocene}. 
Third, the generality of concepts such as modes, iso-mode regions, and evolutionary robustness must be validated across diverse models, and their relevance to social institutions should be further investigated.

In this paper, we introduced an evolutionary dynamical-systems game framework to demonstrate the self-organization of social institutions for common-pool resource management. By coupling players’ actions with environmental dynamics, we examined the evolution of individual decision-making functions that monitor the states of the environment and players as inputs. This approach demonstrated the emergence of norms and punishment, representing institutions.
The game dynamics converge to typical modes of temporally periodic socio-ecological dynamics. We proposed plasticity in decision-making as a necessary property for evolutionary robust institutions. To sum up, this study provides a novel game-theoretic framework for institutional theory, along with general mathematical concepts for games coupling individual decision-making with environmental dynamics.

\section*{Materials and Methods}
\subsection*{Algorithms of the model}
The algorithm for a game step in the model is as follows: For players $1$ and $2$,
\begin{align}
a_1 &= \chi(x(t) + S_1 y_1(t) + O_1 y_2(t)), \label{ds_eq:action1} \\
a_2 &= \chi(x(t) + S_2 y_2(t) + O_2 y_1(t)), \label{ds_eq:action2} \\
r(x(t)) &= 2x(t) - x(t)^2, \label{ds_eq:growth} \\
x(t + 1) &= r(x(t))/3^{(a_1 + a_2)}, \label{ds_eq:resource} \\
p_i(t) &= a_i(x(t + 1) - r(x(t)))/ (a_1 + a_2), \label{ds_eq:harvest} \\
y_i(t + 1) &= 0.8 y_i(t) + p_i. \label{ds_eq:state}
\end{align}
In each step, players decide their actions based on their decision-making functions using the states of the environment and players as inputs (Eqs. (\ref{ds_eq:action1}) and (\ref{ds_eq:action2})). Natural resources grow according to \eqref{ds_eq:growth} and are harvested as described in \eqref{ds_eq:resource}. Finally, players' states are updated based on \eqref{ds_eq:harvest} and \eqref{ds_eq:state}.

At the end of a generation, players' fitnesses are calculated as $h_i = \sum_{t = 0}^{T} y_i(t) / T$. Each player $i$ leaves Poisson($h_i / \ev{h}$) offspring, where $\ev{h}$ denotes the average fitness across the population. Offspring inherit their parent's decision parameters $(S, O)$, with small random noise added, having a mean of $0$ and variance $\mu^2$. The population in the next generation is then randomly matched into pairs and allocated to new resource sites.
\subsection*{Acknowledgement}
The authors thank Eizo Akiyama, Hisashi Otsuki, Yoshiya Matsubara, Wataru Toyokawa, Takehiro Tottori, Natsuki Ogusu, and Kim Sneppen for the stimulating discussions.
This research was supported by Special Postdoctoral Researcher Program in RIKEN Project code 202401061006 (KI); and Novo Nordisk Fonden Grant number NNF21OC0065542 (KK).



\bibliographystyle{unsrt}

\section*{Supplementary Figures}
\supplementaryfigures
\supplementarytables
\begin{figure}[H]
  \centering
   \includegraphics[width=\linewidth]{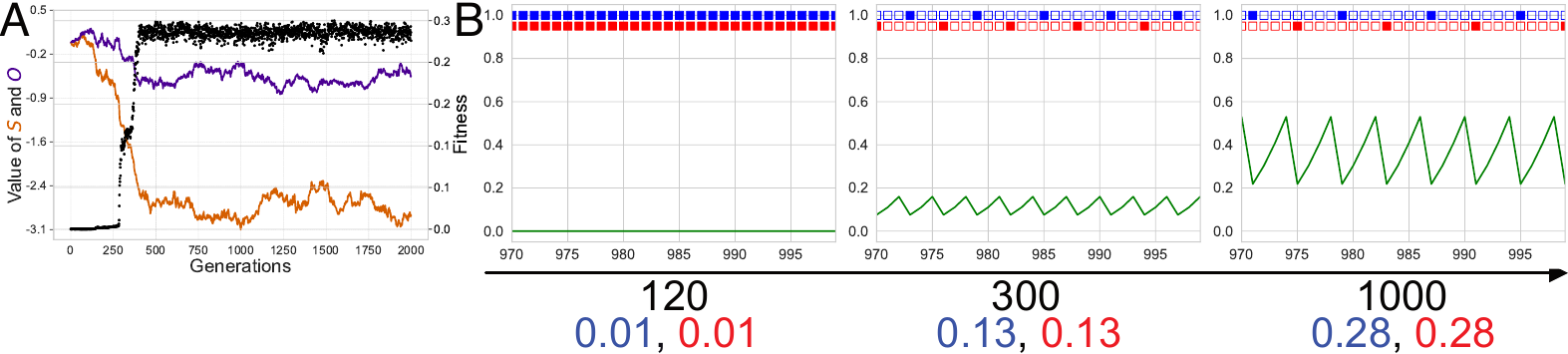}
    \caption{Example of the evolutionary dynamical-systems game where the resource growth rate $\alpha = 0.5$. (A) Generational changes in the average values of the decision parameters $S$ (orange) and $O$ (purple). Black dots indicate the average fitness values (right axis). (B) Examples of game dynamics during the last 30 game steps across several generations. In each panel, blue (and red) boxes represent the harvesting actions of players 1 (and 2), while green represents the resource amount. The numbers below each panel indicate the generation (black) and the fitness values of players 1 (blue) and 2 (red). Parameters are set to $N = 300$ and $\mu = 0.03$.}
    \label{fig:DS_game_temporal_alpha}
\end{figure}

\begin{figure}[H]
  \centering
   \includegraphics[width=\linewidth]{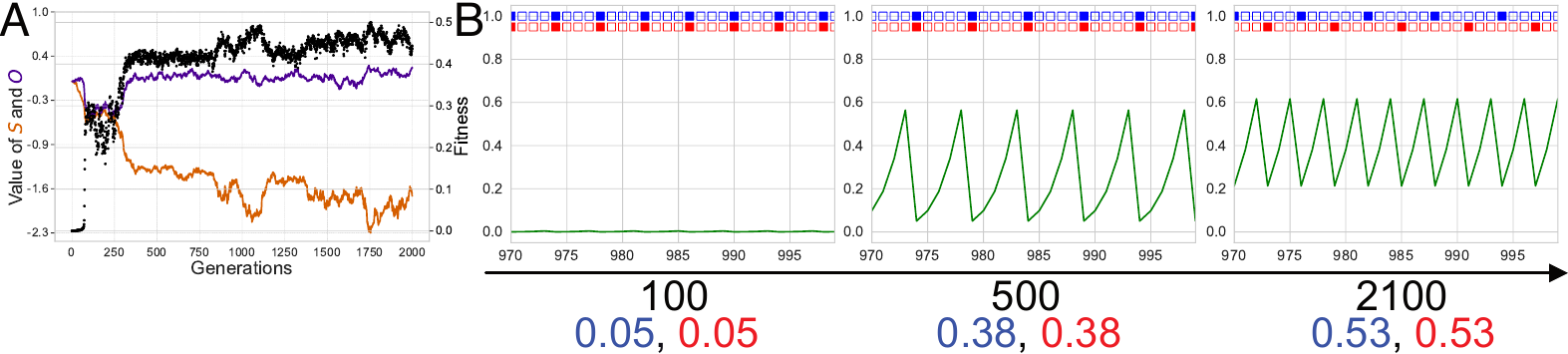}
   \caption{Example of the evolutionary dynamical-systems game where the harvesting ratio $\beta = 3/4$. (A) Generational changes in the average values of the decision parameters $S$ (orange) and $O$ (purple). Black dots indicate the average fitness values (right axis). (B) Examples of game dynamics during the last 30 game steps across several generations. In each panel, blue (and red) boxes represent the harvesting actions of players 1 (and 2), while green represents the resource amount. The numbers below each panel indicate the generation (black) and the fitness values of players 1 (blue) and 2 (red). Parameters are set to $N = 300$ and $\mu = 0.03$.}
    \label{fig:DS_game_temporal_beta}
\end{figure}

\begin{figure}[H]
  \centering
   \includegraphics[width=\linewidth]{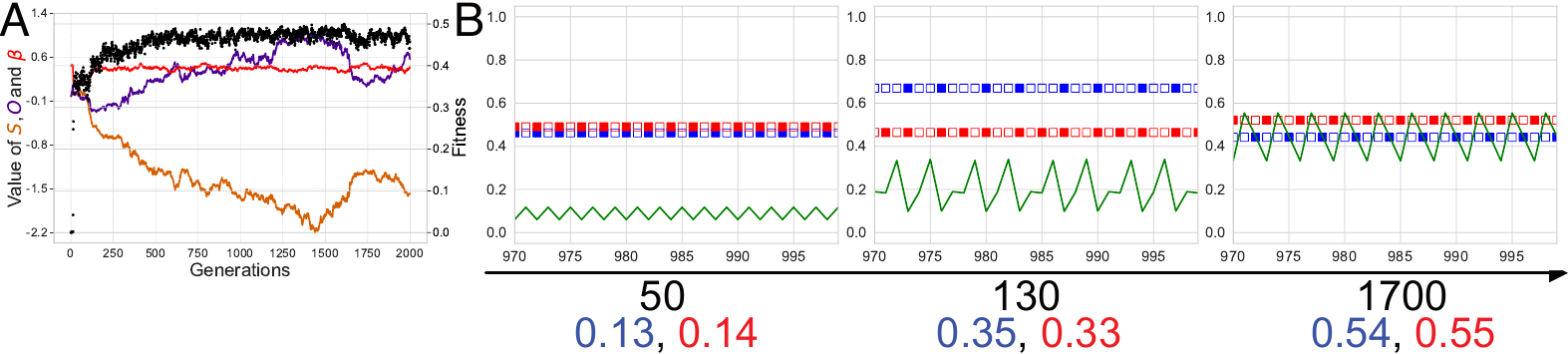}
   \caption{Example of the evolutionary dynamical-systems game where the harvesting ratio $\beta$ evolves over generations. In this example, each harvest incurs a cost of $c = 0.01$. Without such a cost, players tend to harvest small amounts in every step, making the evolution of periodic modes unlikely. Introducing a harvest cost is reasonable, as harvesting actions generally require some expenditure. (A) Generational changes in the average values of the decision parameters $S$ (orange), $O$ (purple), and $\beta$ (red). Black dots indicate the average fitness values (right axis). (B) Examples of game dynamics during the last 30 game steps across several generations. In each panel, blue (and red) boxes represent the harvesting actions of players 1 (and 2), while green represents the resource amount. The levels of boxes indicate the values of $\beta$.
   The numbers below each panel indicate the generation (black) and the fitness values of players 1 (blue) and 2 (red). Parameters are set to $N = 300$ and $\mu = 0.03$.}
    \label{fig:DS_game_temporal_beta_evolve}
\end{figure}

\begin{figure}[H]
  \centering
   \includegraphics[width=\linewidth]{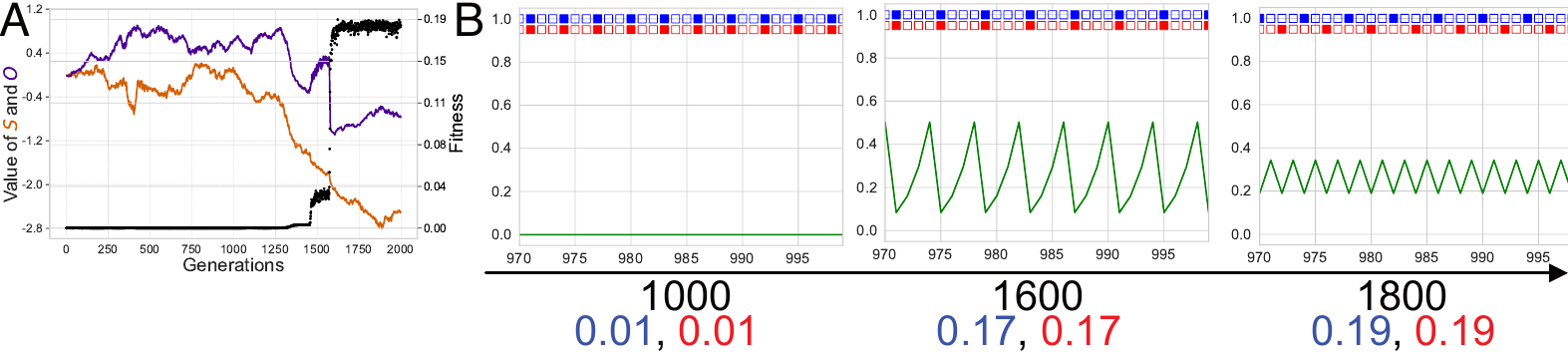}
   \caption{Example of the evolutionary dynamical-systems game where the decay rate of richness $\kappa = 0.5$. (A) Generational changes in the average values of the decision parameters $S$ (orange) and $O$ (purple). Black dots indicate the average fitness values (right axis). (B) Examples of game dynamics during the last 30 game steps across several generations. In each panel, blue (and red) boxes represent the harvesting actions of players 1 (and 2), while green represents the resource amount. The numbers below each panel indicate the generation (black) and the fitness values of players 1 (blue) and 2 (red). Parameters are set to $N = 300$ and $\mu = 0.03$.}
    \label{fig:DS_game_temporal_kappa}
\end{figure}

\begin{figure}[H]
  \centering
   \includegraphics[width=\linewidth]{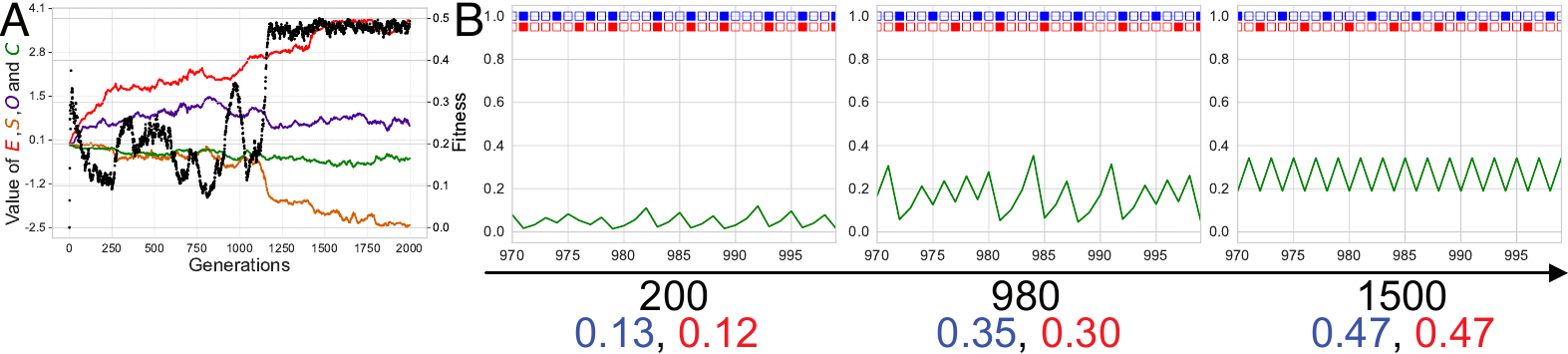}
   \caption{Example of the evolutionary dynamical-systems game with the ``full model'' of decision-making function as $f_i: \chi(E_ix(t) + S_i y_\text{self}(t) + O_i y_\text{opponent}(t) + C_i)$.
   (A) Generational changes in the average values of the decision parameters $E$ (red), $S$ (orange), $O$ (purple), and $C$ (green). Black dots indicate the average fitness values (right axis). (B) Examples of game dynamics during the last 30 game steps across several generations. In each panel, blue (and red) boxes represent the harvesting actions of players 1 (and 2), while green represents the resource amount. The numbers below each panel indicate the generation (black) and the fitness values of players 1 (blue) and 2 (red). Parameters are set to $N = 300$ and $\mu = 0.03$.}
    \label{fig:DS_game_temporal_full}
\end{figure}

\begin{figure}[H]
  \centering
   \includegraphics[width=\linewidth]{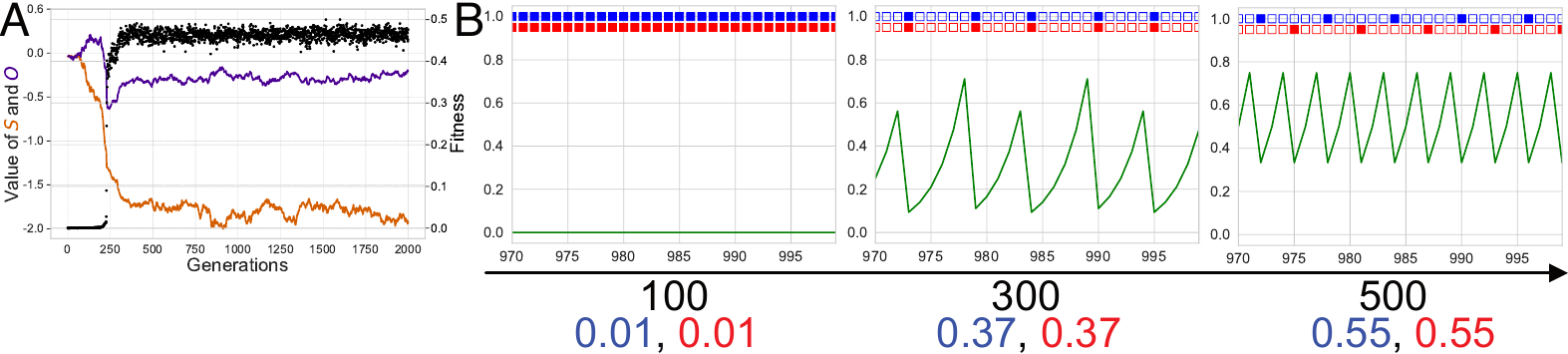}
   \caption{Example of the evolutionary dynamical-systems game where growth function $r(x) = \min(\alpha x, 1.0),$ with $\alpha = 1.5$. (A) Generational changes in the average values of the decision parameters $S$ (orange) and $O$ (purple). Black dots indicate the average fitness values (right axis). (B) Examples of game dynamics during the last 30 game steps across several generations. In each panel, blue (and red) boxes represent the harvesting actions of players 1 (and 2), while green represents the resource amount. The numbers below each panel indicate the generation (black) and the fitness values of players 1 (blue) and 2 (red). Parameters are set to $N = 300$ and $\mu = 0.03$.}
    \label{fig:DS_game_temporal_growth}
\end{figure}

\begin{figure}[H]
  \centering
   \includegraphics[width=\linewidth]{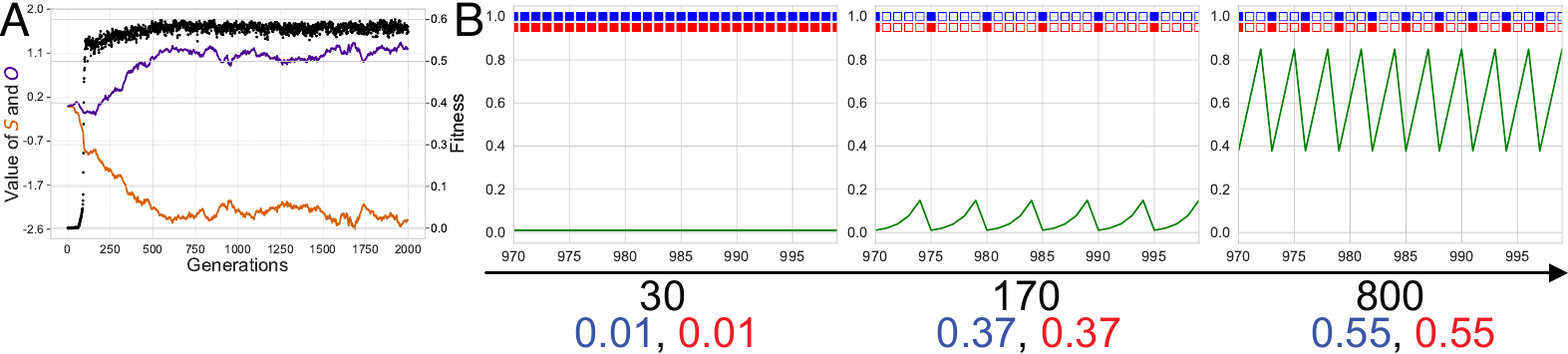}
   \caption{Example of the evolutionary dynamical-systems game where harvest function $x(t + 1) = r(x(t)) - \beta(a_1 + a_2),$ with $\beta = 0.3$. (A) Generational changes in the average values of the decision parameters $S$ (orange) and $O$ (purple). Black dots indicate the average fitness values (right axis). (B) Examples of game dynamics during the last 30 game steps across several generations. In each panel, blue (and red) boxes represent the harvesting actions of players 1 (and 2), while green represents the resource amount. The numbers below each panel indicate the generation (black) and the fitness values of players 1 (blue) and 2 (red). Parameters are set to $N = 300$ and $\mu = 0.03$.}
    \label{fig:DS_game_temporal_harvest}
\end{figure}

\begin{figure}[htb]
  \centering
   \includegraphics[width=\linewidth]{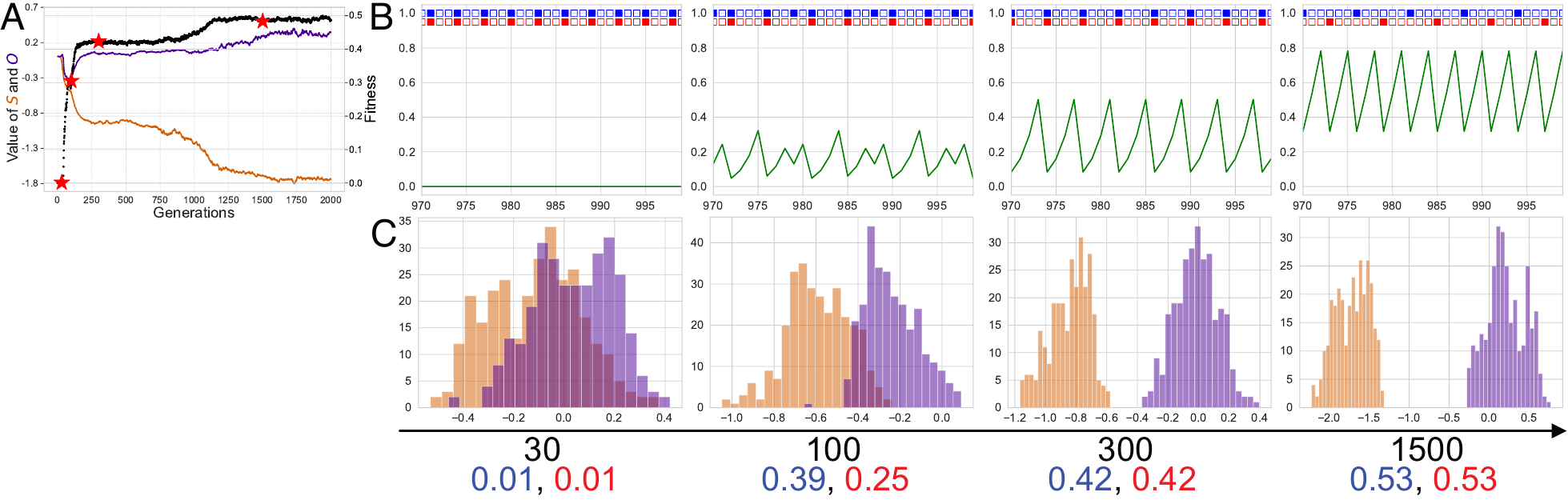}
   \caption{Example of the evolutionary dynamical-systems game. (A) Generational changes in the average values of the decision parameters $S$ (orange) and $O$ (purple). Black dots indicate the average fitness values (right axis). Red stars indicate the samples in (B). (B) Examples of game dynamics during the last 30 game steps across several generations. In each panel, blue (and red) boxes represent the harvesting actions of players 1 (and 2), while green represents the resource amount. The numbers below each panel indicate the generation (black) and the fitness values of players 1 (blue) and 2 (red). (C) Distribution of decision parameters $S$ (red) and $O$ (green) in the system at each generation. Parameters are set $N = 300$ and $\mu = 0.03$.}
    \label{fig:DS_game_temporal_original_si}
\end{figure}

\begin{figure}[htb]
  \centering
   \includegraphics[width=.5\linewidth]{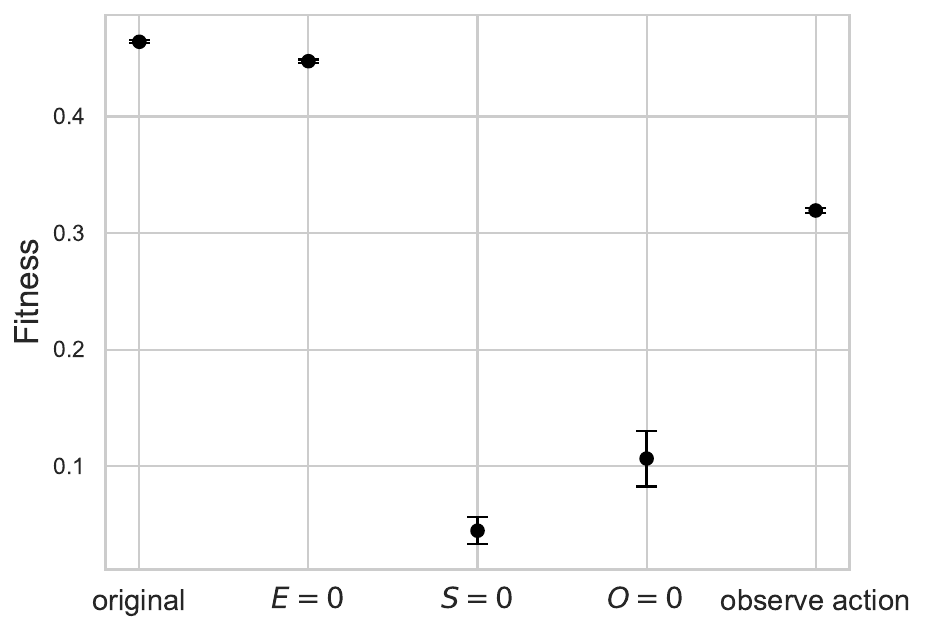}
    \caption{The average fitness for several model variants. The graph shows the average values of fitness for the model variants (original, $E\equiv 0$, $S\equiv 0$, $O\equiv 0$, and observing players' actions instead of states). Bars show the standard deviations. Parameters are $N = 100$ and $\mu = 0.03$. In the variants with either $E, S$ or $O\equiv 0$, one parameter in the decision-making function $f_i = \chi(E_ix(t) + S_i y_\text{self}(t) + O_i y_\text{opponent}(t))$ is fixed to $0$ throughout the evolution.
}
    \label{fig:DS_game_fitness_bar}
\end{figure}
\clearpage
\begin{figure}[H]
  \centering
   \includegraphics[width=\linewidth]{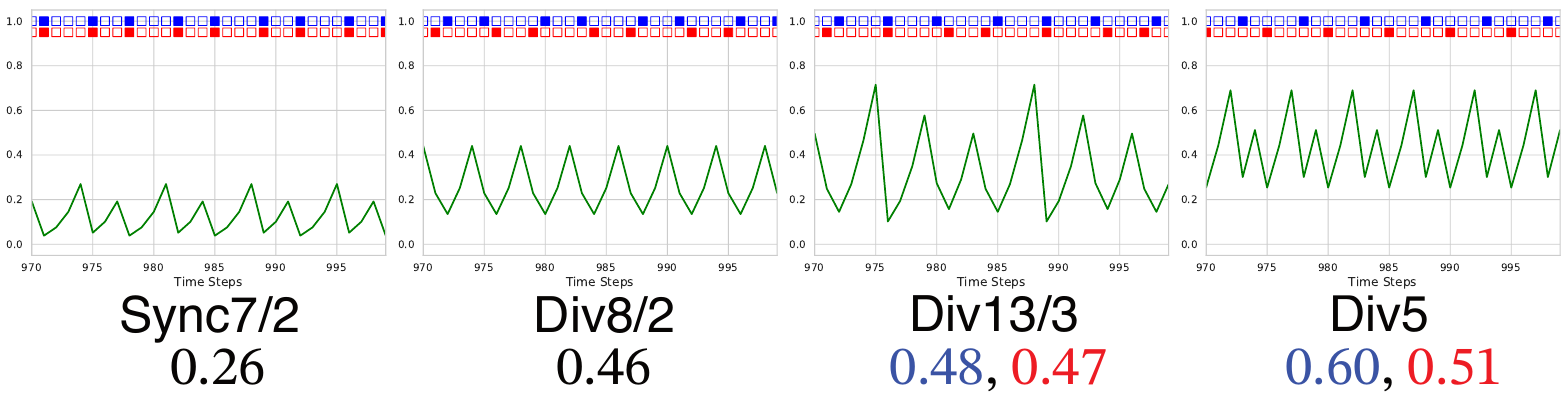}
    \caption{The variety of modes.  Modes are classified based on the period and synchronicity of harvesting actions. ``Sync'' indicates that the actions of the two players are synchronized, and ``Div'' indicates that they are temporarily divided. The subsequent number denotes the period of the players' actions divided by the number of harvesting actions during that period. In each panel, blue (and red) boxes represent the harvesting actions of players 1 (and 2). Green indicates the resource amount. The values at the bottom are fitness. Fitness is shown in black, if identical. Otherwise, fitness in blue (and red) shows that of player $1$ (and $2$).}
    \label{fig:DS_game_phase_examples_si}
\end{figure}

\begin{figure}[H]
  \centering
   \includegraphics[width=.8\linewidth]{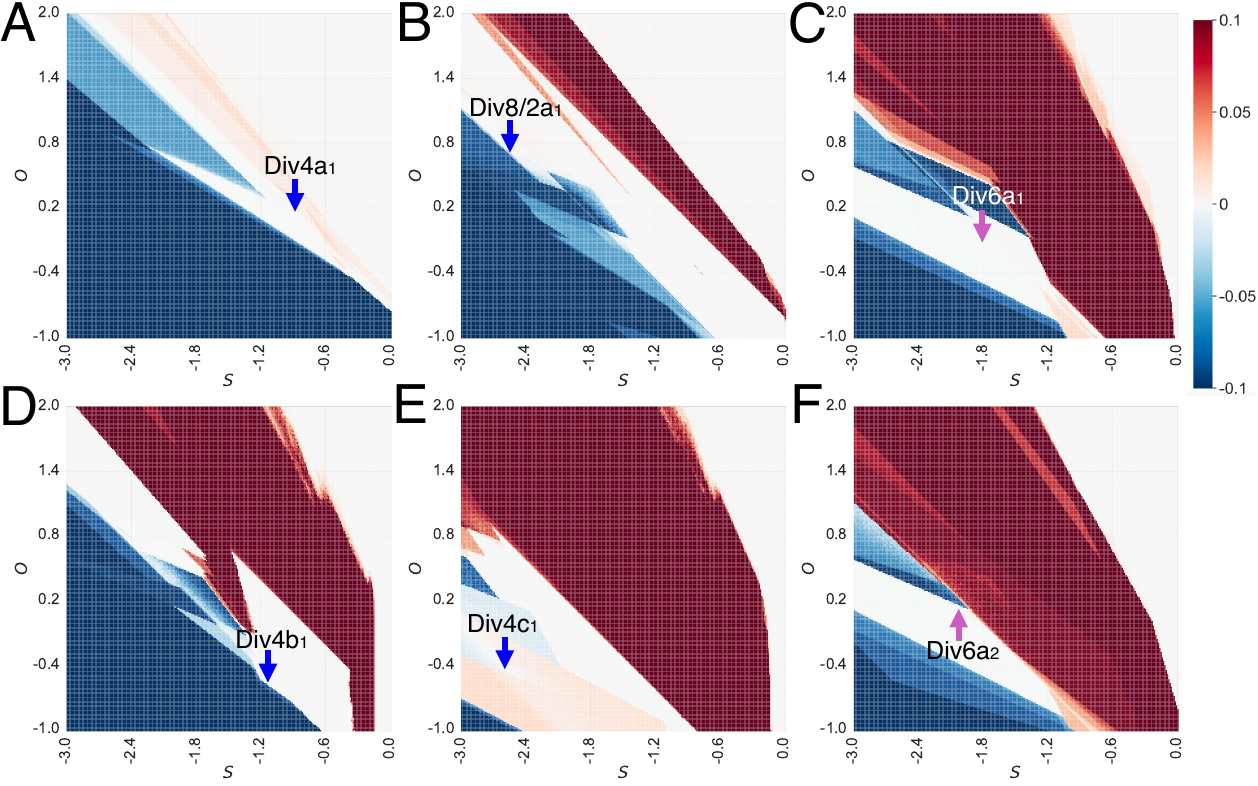}
    \caption{Evolutionary invadability of strategies (Div4$a_1$ (A), Div8/2$a_1$ (B), Div6$a_1$ (C), Div4$b_1$ (D), Div4$c_1$ (E), and Div6$a_2$ (F), named as ``[mode name][region ID][strategy number]''). One strategy is fixed at the arrowhead and the opponent's strategy is varied in the 2D space. One strategy is fixed at the arrowhead while the opponent's strategy is varied in 2D parameter space. Colors indicate the fitness of the opponent as the difference from that of the original strategy (i.e., $h(f, f^\ast) -h(f^\ast, f)$). The fixed strategy can invade the strategies plotted in blue.}
    \label{fig:DS_game_path_si}
    \end{figure}

\clearpage

\section*{Supplementary Text}
\subsection*{Proof of Proposition 1}
\begin{prop}
    Modes depend solely on the action sequences $\{(a_1(i), a_2(i))\}_{i=1}^{l}$ (or their periods and synchronicity). Thus, modes remain invariant under changes in decision parameters, provided the action sequences are unchanged. For any given periodic action sequence, at most one mode exists.
\end{prop}
\vspace{10pt}

\textit{Proof.} Now, the action sequence is given. Let us consider the case where players synchronically harvest with period $l$. Since $0 < x < 1$,
\begin{align}
    \dv{x}r^l(x) &= \prod_{j = 0}^{l-1} (2 - 2r^j(x)) > 0\\
    \dv[2]{x}r^l(x) &= -2\sum_{j = 0}^{l-1} \dv{x}r^j(x)\prod_{k \neq j}^{l-1} (2 - 2r^k(x)) < 0,
\end{align}
where $r^l(x)$ denotes the $l$-times iteration of $r(x)$.
Thus, $r^l(x)$ is a monotonically increasing concave function of $x$. If $x^\ast(l,\ \text{Sync})$ > 0 exists which satisfies $\dv{x}r^l(x^\ast(l,\ \text{Sync})) = 9$ to cancel the reduction by harvest, such $x^\ast(l,\ \text{Sync})$ is unique for $l$ since $\dv[2]{x}r^l(x) < 0$. As $\dv{x}r^l(0) = 2^l$, $x^\ast(l,\ \text{Sync})$ exists for $l > \log_2 9 \simeq 3.17$. If $x$ after harvest is smaller than $x^\ast(l,\ \text{Sync})$, $r^l(x) / 9 > x$, whereas if  $x$ after harvest is larger than $x^\ast(l,\ \text{Sync})$, $r^l(x) / 9 < x$. Hence, with time, $x$ after harvest converges to $x^\ast(l,\ \text{Sync}) = r^l(x^\ast(l,\ \text{Sync})) / 9$, and the periodic change in the states of the environment and players converges to the identical limit cycle. Limit cycles are stable and chaos never appears. Thus, for each period $l$, there is only one limit cycle, characterized by the resource amount after harvest $x^\ast(l,\ \text{Sync})= r^l(x^\ast(l,\ \text{Sync})) / 9$.

For the complex action sequence given by $\{(a_1(i), a_2(i))\}_{i=1}^{l}$, the mode with $\{x(i)\}_{i=1}^{l}$ satisfying $x(1) = \allowbreak r(x(l)) / 3^{a_1(l) + a_2(l)}$ and $x(i + 1) = r(x(i))/ 3^{a_1(i) + a_2(i)}$ for $1 \le i \le l-1$ emerge. With a similar analysis to the above, it is proven that, for given periodic action sequences, at most one mode exists. (proof ends.)

\subsection*{Analytical calculation of the model}
Here, we analytically discuss the properties of modes. First, we discuss the conditions for general nonlinear growth rate $\alpha$ and harvesting fraction $\beta$ where the above Prop.1 holds. In the latter part, we specifically analyze the current conditions with $\alpha = 1$ and $\beta = 2 / 3$.


\begin{figure}[b]
  \centering
   \includegraphics[width=\linewidth]{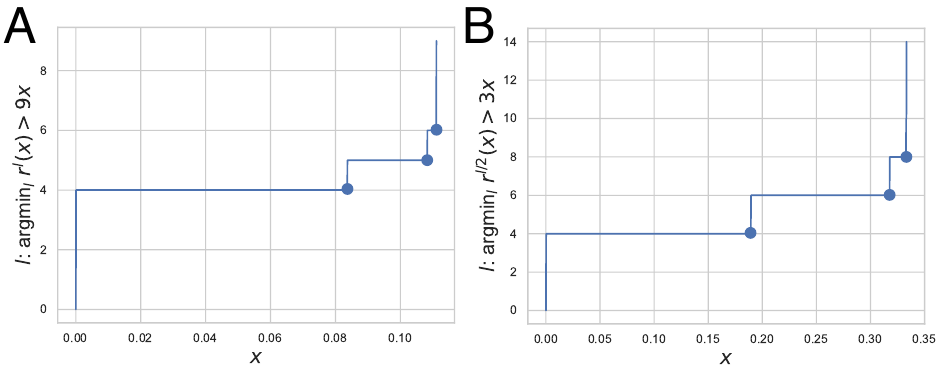}
   \caption{The minimal cycle length $l(x)$ required for sustainability. For a given resource level $x$, if players harvest the resource before $l(x)$ steps have passed, the resource level will eventually decline to $0$. 
   (A) Minimal cycle length $l(x)$ for synchronized actions. 
   (B) Minimal cycle length $l(x)$ for labor division. In both cases, the resource level after harvesting converges to the blue dots, which satisfy $\max_{x} r^l(x) > 9x$ or $\max_{x} r^{l/2}(x) > 3x$.}
    \label{fig:DS_game_analytic}
\end{figure}

First, the above Prop.1 holds for any harvesting fraction $\beta$ ($0 < \beta < 1$) if $r^l(x)$ is a monotonically increasing concave function of $x$.
When the resource growth is given by $r(x) = x + \alpha (x - x^2)$, the resource amount $x$ varies within $0 < x < 1$ if $\alpha \le 1$ and within $0 < x < (1 + \alpha)^2 / 4a$ if $\alpha > 1$. If $\alpha \le 1$, $r^l(x)$ is a monotonically increasing concave function of $x$, and thus the above proof holds. By contrast, if $\alpha > 1$, $r(x)$ monotonically increases in $0 < x < (1 + \alpha) / 2$ and decreases in $(1 + \alpha) / 2 < x < (1 + \alpha)^2 / 4a$. The above proof does not generally hold. Still, as the evoved decision-making function will realize sustainable and efficient resource use, it is expected that the resource amount $x$ varies within $0 < x < (1 + \alpha) / 2$ and the above proof will effectively hold.

Let us focus on the current situation with $\alpha = 1$ and $\beta = 2 / 3$.
When harvesting in synchrony, the resource amount becomes $1 / 9$, and when dividing labor, it becomes $1 / 3$. If each is done every $l$ cycle, the resource amount $x_0$ immediately after harvesting converges to $x^\ast = r^l(x^\ast) / 9$ in the case of synchronization and $x^\ast = r^{l /2}(x^\ast) / 3$ in the case of labor division.

In the case of synchronization, since $r(x) = 2x - x^2 < 2x$, we have $r^3(x) < 8x < 9x$. Therefore, with $l \le 3$, $x$ converges to $0$. To maintain resource sustainability, the cycle must be at least 4 (although a mixed strategy of 3 and 4 cycles could be feasible). 
For a cycle of 4, the resource amount after harvesting asymptotically approaches $x^\ast = 0.084$, as shown in Fig. \ref{fig:DS_game_analytic}. At the time of harvesting, $r^4(x^\ast) = 0.754$, and $0.336$ is harvested each time.
Considering the steady state of the resource, we solve $y(t_{n + 1}) = 0.8^4 y(t_n) + 0.336 = y(t_n)$ and obtain $y(t_n) = 0.57$, where $t_n$ denotes the time of $n$th harvesting action. Thus, the fitness is given by $h = y(t_n) (1 + 0.8 + 0.8^2 + 0.8^3) / 4 = 0.42$. Similarly, we can calculate the properties of each mode. Although the cycle with $l = 8$ is possible in Fig. \ref{fig:DS_game_analytic}, it exhibits lower fitness $0.42$ due to too much waiting and thus does not evolve in the simulation.


\begin{table}[tb]
    \centering
    \caption{Properties of each mode. ``H'' indicates the moment when a player decides to harvest, and ``W'' indicates the moment of deciding to wait just before harvesting. In the Sync mode, the states of both players are equal. By contrast, in the Div mode, the states of the players differ, with the opponent's state shown in parentheses.}
    \begin{tabular}{c|ccccc}
       mode  & fitness & $x$ (at H) & $y$ (at H) & $x$ (at W) & $y$ (at W) \\\hline
       Sync4  & 0.42 & 0.50 & 0.27 & 0.30 & 0.34 \\
       Sync5 & 0.43 & 0.84 & 0.33 & 0.60 & 0.41\\
       Div4 & 0.47 & 0.34 & 0.33 (0.64) & 0.19 & 0.41 (0.51)\\
       Div6 & 0.53 & 0.78 & 0.28 (0.55) & 0.53 & 0.35 (0.69)\\
       Div8 & 0.42 & 0.96 & 0.17 (0.41) & 0.80 & 0.21 (0.51)
    \end{tabular}
    
    \label{tab:DS_game_calc}
\end{table}

Table \ref{tab:DS_game_calc} summarizes the amount of resources and the states of players at the timing of decision-making, and resulting fitness for each mode.
To realize Sync4, for instance, $x(\tau) + Sy_1(\tau) + Oy_2(\tau)$ should be positive for $\tau = 4$ (at harvesting) and negative for $\tau = 1, 2, 3$ (at waiting). As $x$ monotonically decreases and $y_1$ and $y_2$ monotonically increase for $\tau = 1, 2, 3$, we only have to consider the cases for $\tau = 3$ and $4$.
According to Table \ref{tab:DS_game_calc}, these conditions are $0.3 + 0.34(S + O) < 0$ and $0.5 + 0.27(S + O) > 0$. Hence, Sync4 can be realized if $-1.85 < S+0 < -0.88$. 

By contrast, to realize Div4, for player 1, $x(\tau) + Sy_1(\tau) + Oy_2(\tau)$ should be positive only for $\tau = 4$ (at player 1 harvesting) and negative for $\tau = 1, 3$ (at both waiting) and for $\tau = 2$ (at player 2 harvesting). Hence, $S$ and $O$ should satisfy $0.34 + 0.33S + 0.64O > 0$, $0.19 + 0.51S + 0.41O < 0$, $0.34 + 0.64S + 0.33O < 0$, and $0.19 + 0.41S + 0.51O < 0$. Here, both $S / O$ and $S + O$ are essential to realize Div4.

Therefore, the conditions for synchronized harvesting concern $S + O$ but those for division of labor concern both $S + O$ and $S / O$. For the labor division to occur, it is necessary to wait while one's state is larger than the opponent's, requiring a certain balance of $S / O$. Division of labor requires the waiting decision when the self-state is richer than the opponent, it is reasonable that Div4 and Div6 evolve when $S << O$. Additionally, in Div6, players wait longer until the resource grows sufficiently (harvesting if $x = 0.78$) than in Div4 (harvesting if $x = 0.34$). Thus, a smaller $S$ is needed to realize Div6.

\end{document}